\documentclass[aip,amsmath,amssymb,reprint,floatfix]{revtex4-2}
\newcommand{\bol}[1]{\boldsymbol{#1}}
\usepackage{balance}
\usepackage{physics}
\usepackage{dcolumn}
\usepackage{calc}
\usepackage{mathtools}
\usepackage{amsfonts}
\usepackage{natbib}
\usepackage{comment}
\usepackage[pdftex,pdfpagelabels,bookmarks,hyperindex,hyperfigures,colorlinks]{hyperref}
\draft 
\usepackage[caption=false]{subfig}
\begin{document}


\title{Quantum Computing Perspective for Electromagnetic Wave Propagation in Cold Magnetized Plasmas} 



\author{Efstratios Koukoutsis}
\email{stkoukoutsis@mail.ntua.gr}
\author{Kyriakos Hizanidis}
\affiliation{School of Electrical and Computer Engineering, National Technical University of Athens, Zographou 15780, Greece}
\author{George Vahala}
\affiliation{Department of Physics, William \& Mary, Williamsburg, Virginia 23187, USA}
\author{Min Soe}
\affiliation{Department of Mathematics and Physical Sciences, Rogers State University, Claremore, Oklahoma 74017, USA}
\author{Linda Vahala}
\affiliation{Department of Electrical and Computer Engineering, Old Dominion University, Norfolk, Virginia 23529, USA}
\author{Abhay K. Ram}
\affiliation{Plasma Science and Fusion Center, Massachusetts Institute of Technology, Cambridge, Massachusetts 02139, USA}

\date{\today}

\begin{abstract}
Electromagnetic waves are an inherent part of all plasmas – laboratory fusion plasmas or astrophysical plasmas. The conventional methods for studying properties of electromagnetic waves rely on discretization of Maxwell equations suitable for implementing on classical, present day, computers. The traditional methodology is not efficient for quantum computing implementation – a future computational source offering a tantalizing possibility of enormous speed up and a significant reduction in computational cost. This paper addresses two topics relevant to implementing Maxwell equations on a quantum computer. The first is on formulating a quantum Schrödinger representation of Maxwell equations for wave propagation in a cold, inhomogeneous, magnetized plasma. This representation admits unitary, energy preserving, evolution and conveniently lends itself to appropriate discretization for a quantum computer. Riding on the coattails of these results, the second topic is on developing a sequence of  unitary operators which form the basis for a qubit lattice algorithm (QLA). The QLA, suitable for quantum computers, can be implemented and tested on existing classical computers for accuracy as well as scaling of computational time with the number of available processors. In order to illustrate the QLA for Maxwell equations, results are presented from a time evolving, full wave simulation of propagation and scattering of an electromagnetic wave packet by non-dispersive dielectric medium localized in space.
\end{abstract}

\maketitle 


\section{Introduction}\label{sec:1}
Propagation of electromagnetic waves in thermonuclear fusion plasmas is one of the most significant fields of research in the pursuit for magnetic
fusion. In magnetic confinement experiments, electromagnetic waves play a vital role in plasma temperature control, localized non-inductive current drive, heating, and  plasma instability control. Therefore, there is an utmost need for understanding the physics and mechanics of wave propagation and scattering inside an inhomogeneous magnetized plasma to enable the optimization for fusion applications.

While the bedrock for the theoretical and analytical studies of wave propagation in plasmas has long been established,\cite{Stix,Swanson} penetrating into the complex processes that occur in  plasmas and unraveling their physics require a computational treatment. To that end, taking into consideration the aforementioned importance of electromagnetic wave propagation in plasmas, a plethora of computational tools have been developed,\cite{Friedland,Tsironis,Lau} ranging from ray-tracing methods to full-wave simulations along with different domains of application.

However, solving the mathematical and physical problem of wave propagation in an actual fusion device poses a challenge even for the most advanced supercomputers. With classical computers eventually reaching their limits and fusion research heavily relying on computational results we motivate a shift in the traditional computational methods, engaging the modern and uprising quantum technologies and quantum computing in particular.

Quantum computing is one of those computational pathways that can yield faster computations than those achieved on a classical computer, \cite{Wu,Arute} the so called quantum advantage, and has gained significant attention in the plasma physics community. Considerations on general applications in plasma simulation  can be found in Ref.[\onlinecite{Dodin}], whereas a fusion oriented review of possible quantum computing applications is Ref.[\onlinecite{Joseph}]. In Refs. [\onlinecite{Engel}] and [\onlinecite{Novikau}] the authors  exploit the Quantum Signal Processing (QSP) protocol \cite{Low}  for simulation of electrostatic Landau damping and wave propagation in a cold fluid plasma respectively. In addition, a quantum computing treatment for Vlasov equation with collisions has been presented in Ref. [\onlinecite{Ameri}]. Finally, a comprehensive review on quantum computing applications in plasmas can be found in Ref.[\onlinecite{Amaro}]. 

In this paper, we examine Maxwell equations for wave propagation in cold, inhomogeneous, magnetized plasma amendable to quantum computing without tackling the question of computational advantage over the classical methods. Quantum computers are restricted to unitary operations following the physical laws of closed quantum systems. Thus, the first step towards a quantum implementation is to reformulate Maxwell equations as a quantum Schrodinger equation with Hermitian structure, extending the results of [\onlinecite{Koukoutsis}] to encompass the  dispersive nature of cold magnetized plasma. Then, the second challenge  entails decomposing the relevant unitary operator of evolution into a product sequence of unitary operators that can be encoded efficiently on a quantum computer. We accomplish this by leveraging  the tensor product structure of the Hamiltonian, deriving a Trotterized unitary sequence that constitutes the basis for a latter Qubit Lattice Algorithm (QLA). The scaling of the quantum encoded QLA has been recently reported\cite{Succi,Gourdeau,Koukoutsis} to favor quantum implementation on real quantum hardware. 

Qubit lattice algorithms along with its predecessors have found extensive computational applications in the fields of Maxwell equations,\cite{Vahala1,Vahala2,Ram,Vahala3,Vahala4,unpublished} non-linear optics\cite{Vahala5,Linda1} and quantum simulations.\cite{Yepez,Yepez2,Boghosian,Koukoutsis}

To assess the capabilities of QLA we present full-wave simulation results from propagation and scattering of an electromagnetic wave packet in a reduced case of our formulation,\cite{unpublished} for a localized inhomogeneous, scalar dielectric. Such wave packet structures in plasma are related to the finite spatial extent applied RF waves that are routinely used for plasma heating. Although these simulations are implemented on classical supercomputers they can be directly transferred to quantum computers, acting as a precursor and validation factor for the proposed QLA generalization into cold magnetized plasma in the near term future. 

This paper is structured in two main sections. Section \ref{sec:2} sets up the theoretical formulation of Maxwell equations as a quantum Schrodinger equations, following up a decomposition of the evolution operator into a convenient unitary product sequence for QLA discretization along with the pertinent discussion on complexity. In Sec.\ref{sec:2a} an augmented form of Maxwell equations in magnetized plasma is presented, serving as a stepping stone for the construction of a Schrodinger-Maxwell equation with unitary evolution in Sec.\ref{sec:2b}. The importance of initial and boundary conditions is discussed in Sec. \ref{sec:2c}. Decomposition of the established unitary evolution in a product formula of simple unitary operators based on Trotterization is the main subject of Sec.\ref{sec:2d}. A simple complexity analysis is performed in Sec.\ref{sec:2e} regarding the scaling of QLA implementation in quantum hardware, indicating a polynomial scaling with the number of qubits required for the QLA discetization. Then, a commentary section \ref{sec:2f} follows, containing perspectives on the QLA implementation for wave propagation and scattering in the cold plasma. Section \ref{sec:3} serves as an indicator of QLA capabilities for the future implementation in the cold plasma case studied in Sec.\ref{sec:2}. Specifically, in sections \ref{sec:3a} and \ref{sec:3b} we present the algorithmic scheme of QLA along with some initial value simulations for full-wave scattering of an electromagnetic wave-packet from two-dimensional (2D) scalar, non-dispersive inhomogeneous dielectric objects. In particular, we contrast the different scattering characteristics from a local cylindrical dielectric with strong gradients in the finite boundary layer between the dielectric and vacuum, with that scattering from a local conic dielectric with weak boundary layer gradients in the refractive index. Finally, in Sec.\ref{sec:4} we discuss our results along with the next necessary steps for an actual QLA implementation in the near future.

\section{Quantum implementation of Maxwell equations in cold magnetized plasma}\label{sec:2}
For a non-dispersive, tensorial and inhomgeneous medium, Maxwell equations can be written as a Schrodinger equation with unitary evolution\cite{Koukoutsis}
\begin{equation}\label{1}
i\pdv{\bol{\psi}}{t}=\hat{D}_\rho\bol\psi,\quad\hat{D}_\rho=\hat{D}^\dagger_\rho,\quad \bol\psi(\bol{r},0)=\bol\psi_0,
\end{equation}
under a Dyson transformation $\hat\rho$ on the electromagnetic fields $\bol{u}=(\bol{E}, \bol{H})^T$, with $\bol\psi=\hat{\rho}\bol{u}$. 
In particular, the Hermitian operator $\hat{D}_\rho$ 
\begin{equation}\label{2}
\hat{D}_\rho=\hat\rho\hat{D}\hat{\rho}^{-1}=\hat\rho\hat{W}^{-1}(\bol{r})\hat{M}\hat{\rho}^{-1} ,
\end{equation}
with
\begin{equation}\label{3}
\hat{M}=i\begin{bmatrix}
0_{3\times3}& \bol\nabla\times\\
- \bol\nabla\times & 0_{3\times3}
\end{bmatrix},\quad\hat{W}=\begin{bmatrix}
\epsilon(\bol{r}) &  0_{3\times3}\\
 0_{3\times3} & \mu(\bol{r})
\end{bmatrix}.
\end{equation}
In Eq.\eqref{3} the $\hat{M}$ operator is the Maxwell curl operator and the Hermitian, positive definite $\hat{W}$ matrix represents the constitutive relations of the medium. The explicit form of the Dyson map $\hat{\rho}$ depends on the structure of the material matrix $\hat{W}$: $\hat{\rho} = \sqrt{\hat{W}}$.

On the other hand, the cold magnetized plasma as a dielectric medium is characterized by dispersion. This translates into a frequency dependent permittivity matrix $\Tilde{\epsilon}(\omega)$. Following the Stix notation\cite{Stix},
\begin{equation}\label{13}
\Tilde{\epsilon}(\omega)=\begin{bmatrix}
S&-iD&0\\
iD&S&0\\
0&0&P
\end{bmatrix}
\end{equation}
with
\begin{align}\label{14}
S=&\epsilon_0\Big(1-\sum_{j=i,e}\frac{\omega^2_{pj}}{\omega^2-\omega_{cj}^2}\Big) \nonumber\\
D=&\epsilon_0\sum_{j=i,e}\frac{\omega_{cj}\omega^2_{pj}}{\omega(\omega^2-\omega_{cj}^2)}\\
P=&\epsilon_0\Big(1-\sum_{j=i,e}\frac{\omega^2_{pj}}{\omega^2}\Big) \nonumber.
\end{align}

The definition of elements \eqref{14} in the Stix permittivity tensor is taken for a two-species, ions (i) and electrons (e), plasma with inhomogeneous plasma frequency $\omega^2_{pj}(\bf{r})=\frac{n_j(\bf{r})q^2_j}{m_j\epsilon_0}$ and cyclotron frequency $\omega_{cj}=\frac{q_jB_{0}}{m_j}$. The homogeneous magnetic field $B_0$ is along the $z$ axis and $m_j$, $q_j$ are the mass and charge of the $j$-species respectively.
$n_j(\bf{r})$ is the $j^{th}$ species density.

\subsection{Maxwell equations in temporal domain}\label{sec:2a}
In contrast to the optical response case, the temporal domain transformation of $\Tilde{\epsilon}(\omega)$ is expressed through a convolution integral. As a result, the temporal domain, constitutive relations for a cold magnetized plasma are
\begin{equation}\label{15}
\bol d=\hat{W}_0\bol u+\frac{1}{2\pi}\int_0^t\int_{-\infty}^\infty(\Tilde{\epsilon}(\omega)-\epsilon_0I_{3\times3})e^{-i\omega(t-\tau)}\bol E(\bol r, \tau)d\,\omega d\,\tau,
\end{equation}
with $\bol d=(\bol D, \bol B)^T$. The matrix $\hat{W}_0$ represents the optical response, as in Eq.\eqref{3}, but now only that of the vacuum. Evaluation of the inner integral term in Eq. \eqref{15} requires the Plemelj formula\cite{Stix} to yield
\begin{equation}\label{16}
\bol d=\hat{W}_0\bol u+\int_0^t\hat{K}(t-\tau)\bol E(\bol r, \tau)d\,\tau,
\end{equation}
with the inhomogeneous susceptibility kernel $\hat{K}(t)$ 
\begin{equation}\label{17}
\hat{K}(t)=\epsilon_0\sum_{j=i,e}\begin{bmatrix}
\frac{\omega^2_{pj}}{\omega_{cj}}\sin{\omega_{cj}t}&\frac{\omega^2_{pj}}{\omega_{cj}}(\cos{\omega_{cj}t}-1) &0\\
\frac{\omega^2_{pj}}{\omega_{cj}}(1-\cos{\omega_{cj}t})&\frac{\omega^2_{pj}}{\omega_{cj}}\sin{\omega_{cj}t}&0\\
0&0&\omega^2_{pj}t
\end{bmatrix}.
\end{equation}
From the expressions \eqref{16} and \eqref{17}, Maxwell equations for a cold magnetized plasma now take the form
\begin{equation}\label{18}
i\pdv{\bol u}{t}=W_0^{-1}\hat{M}\bol u-i\int_0^t\pdv{\hat{G}(t-\tau)}{t}\bol u(\bol r,\tau)d\,\tau
\end{equation}
where
\begin{widetext}
\begin{equation}\label{19}
\pdv{\hat{G}(t)}{t}=\begin{bmatrix}
\frac{1}{\epsilon_0}\pdv{\hat{K}}{t}&0_{3\times3}\\
0_{3\times3}&0_{3\times3}
\end{bmatrix},\quad\frac{1}{\epsilon_0}\pdv{\hat{K}}{t}=\sum_{j=i,e}\omega^2_{pj}(\bf{r})
\begin{bmatrix}
\cos{\omega_{cj}t}&-\sin{\omega_{cj}t} &0\\
\sin{\omega_{cj}t}&\cos{\omega_{cj}t}&0\\
0&0&1
\end{bmatrix}.
\end{equation}
\end{widetext}

\subsection{Schrodinger representation}\label{sec:2b}
Returning back to $\Tilde{\epsilon}(\omega)$ in Eq. \eqref{13}, its Hermitian structure ensures that the conductivity current does not produce dissipation inside the plasma, i.e the cold magnetized plasma is a lossless dispersive dielectric. Hence, it is possible to construct a Schrodinger representation of Maxwell equations \eqref{18} that admit unitary evolution corresponding to electromagnetic energy conservation. Such mathematical representations of Maxwell equations for lossless dispersive media are well studied in the literature\cite{Silveirinha,Cassier}.

Defining the total conductivity current density $\bol J_c$ as
\begin{equation}\label{20}
\bol J_c=\int_0^t\pdv{\hat{K}}{t}\bol E(\bol r, \tau)d\,\tau=\bol J_{ce}+\bol J_{ci},
\end{equation}
we exploit the rotational symmetry of $\pdv{\hat{K}}{t}$ in Eq.\eqref{19} to reformulate Maxwell equations \eqref{18} as
\begin{align}\label{21}
i\pdv{\bol E}{t}&=\frac{i}{\epsilon_0}\bol\nabla\times\bol H-\frac{i}{\epsilon_0}\bol J_c,\nonumber\\
i\pdv{\bol H}{t}&=-\frac{i}{\mu_0}\bol\nabla\times\bol E,\\
i\pdv{\bol J_{cj}}{t}&=i\epsilon_0\omega^2_{pj}(\bol{r})\bol E+\omega_{cj}\hat{S}_z \bol{J}_{cj},\quad j=i,e.\nonumber
\end{align}
The set of equations \eqref{21} represent the augmented Maxwell system which self-consistently describes the behaviour of electromagnetic fields inside a cold magnetoplasma.
We point out that Eq.\eqref{21} is the basis for FDTD simulations,\cite{Lee} but for a stationary plasma. The Hermitian matrix $\hat{S}_z$,
\begin{equation}\label{22}
\hat{S}_z=\begin{bmatrix}
0&-i&0\\
i&0&0\\
0&0&0
\end{bmatrix}
\end{equation}
represents the projection of spin-1 onto the $z$-axis.

To obtain an explicit Schrodinger representation of Eq.\eqref{21} we  apply a Dyson transformation\cite{Koukoutsis},
\begin{equation}
\hat{\rho}=diag(\epsilon^{1/2}_0I_{3\times3}, \mu^{1/2}_0I_{3\times3}, \frac{1}{\epsilon_0^{1/2}\omega_{pi}}I_{3\times3}, \frac{1}{\epsilon_0^{1/2}\omega_{pe}}I_{3\times3})
\end{equation}
resulting in 
\begin{widetext}
\begin{equation}\label{24}
i\pdv{t}\begin{bmatrix}
\epsilon_0^{1/2}\bol E\\
\mu_0^{1/2}\bol H\\
\frac{1}{e_0^{1/2} \omega_{pi}}\bol J_{ci}\\
\frac{1}{e_0^{1/2} \omega_{pe}}\bol J_{ce}
\end{bmatrix}=\begin{bmatrix}
0_{3\times3}&ic\bol\curl&-i\omega_{pi}&-i\omega_{pe}\\
-ic\bol\curl&0_{3\times3}&0_{3\times3}&0_{3\times3}\\
i\omega_{pi}&0_{3\times3}&\omega_{ci}\hat{S}_z&0_{3\times3}\\
i\omega_{pe}&0_{3\times3}&0_{3\times3}&\omega_{ce}\hat{S}_z
\end{bmatrix}\begin{bmatrix}
\epsilon_0^{1/2}\bol E\\
\mu_0^{1/2}\bol H\\
\frac{1}{e_0^{1/2} \omega_{pi}}\bol J_{ci}\\
\frac{1}{e_0^{1/2} \omega_{pe}}\bol J_{ce}
\end{bmatrix}\Leftrightarrow i\pdv{\bol \psi}{t}=\hat{D}\bol\psi.
\end{equation}
\end{widetext}
It should be noted that we have switched from using the Riemann-Silberstein-Weber\cite{Khan} field representation to the vacuum field representation, and the plasma inhomogeneity is now thrust into the source terms $\bol J_{ci}, \bol J_{ce}$ through the species plasma frequencies $\omega_{pj}(\bf{r})$.
Additionally, Eq.\eqref{24} can be easily extended to incorporate different ions species by adding the respective ion-species current components in the stave vector $\bol\psi$. In realistic fusion experiments there will be hydrogen, deuterium and tritium ions, so their contribution must be included in Eq.\eqref{24} for a complete description  of the total inhomogeneity profiles.

Under suitable Dirichlet boundary conditions the operator $\hat{D}$ in the Schrodinger-Maxwell Eq.\eqref{24} is Hermitian. As a result, the evolution operator $\hat{\mathcal{U}}=e^{-it\hat{D}}$ is unitary and corresponds to the conservation of an extended electromagnetic energy $E(t)$ through the inner product,
\begin{widetext}
    \begin{equation}\label{25}
E(t)=\braket{\bol\psi}=\int_\Omega\Big(\epsilon_0\abs{\bol E}^2+\frac{\abs{\bol B}^2}{\mu_0}\Big)d\,\bol r+\int_\Omega\Big(\frac{\abs{\bol J_{ci}}^2}{\epsilon_0\omega^2_{pi}(\bf{r})}+\frac{\abs{\bol J_{ce}}^2}{\epsilon_0\omega^2_{pe}(\bf{r})}\Big) d\,\bol r=E(0)=\int_\Omega\Big(\epsilon_0\abs{\bol E_0}^2+\frac{\abs{\bol B_0}^2}{\mu_0}\Big)d\,\bol r,\quad\Omega\subset\mathbb{R}^3.
    \end{equation}
\end{widetext}
The extended electromagnetic energy Eq.\eqref{25} consists of two terms. The first term is the standard electromagnetic energy in a vacuum whereas the second term reflects the energy associated with the cold plasma response. We have denoted with $\bol E_0$ and $\bol B_0$ the initial values of the electromagnetic fields. Notice that due to the causality constraint in the plasma response, the initial values of the conductivity currents according to Eq.\eqref{20} are zero,  $\bol J_{ce,i}(t\leq0)=0$.

A subtlety related with the extended electromagnetic energy \eqref{25} is the smoothness of $E(t)$ because of the Laplace Transform in Eq.\eqref{15}. As a result, even for resonant frequencies $\omega=\omega_{cj}$ we obtain a bounded dispersive electromagnetic energy $E_{disp}(t)\leq E(0)$. Thus, it is possible to quantify the resonant energization for each plasma population without considering resonant wave-particle interactions or pertubative approximations for the RF field.

\subsection{Initial and boundary conditions}\label{sec:2c}
In this section we will restate our problem  comparing the imposed mathematical conditions with the ones in a plasma fusion device.

The plasma as a dielectric is considered to be confined inside a volume $\Omega\subset\mathbb{R}^3$ with a boundary surface $\partial\Omega$. By selecting the boundary condition
\begin{equation}\label{26}
\bol n\times\bol E=0,\quad\text{on $\partial\Omega$},
\end{equation}
the ``Hamiltonian operator" $\hat{D}$ in the Maxwell-Schrodinger equation \eqref{24} is Hermitian so the standard quantum-mechanical analogies are present. In fusion devices, the plasma is confined by a vacuum vessel at which the Perfect Electric Conductor (PEC) boundary condition \eqref{26} no longer holds due to electromagnetic losses in the walls. Alteration of the PEC boundary condition results in the non-Hermiticity of the operator $\hat{D}$ and subsequently, a break in the  unitary evolution. In this case, the quantum simulation of the dynamics becomes troublesome. A solution has been proposed in Ref.[\onlinecite{Novikau2}] where instead of the quantum simulation of the Maxwell dynamics, the linear system of equations is solved through quantum singular value decomposition as a boundary value problem.  This approach could run into some difficulties as one moves to 2D and 3D plasma wave propagation.  
Alternatively, one could resort to some dilation by embedding the subsystem into a higher dimensional Hilbert space and thereby recover unitarity within this higher dimensional space.

For completeness, one could eventually introduce into the set of equations \eqref{21} the effect of an antenna by coupling the Faraday equation with a monochromatic oscillator\cite{Novikau} $\bol Q(\bol r, t)=\bol Q_a(\bol r_a)e^{-i\omega_a t}$ with frequency $\omega_a$. The subscript $a$ denotes the antenna-related quantities. In that way, the Faraday equation in \eqref{24} is augmented by 
\begin{equation}\label{27}
    \begin{aligned}
    i\pdv{(\mu_0^{1/2}\bol H)}{t}&=-ic\bol\curl(\epsilon_0^{1/2}\bol E)+\beta_{\bol r,\bol r_a}\bol Q\\
    i\pdv{\bol Q}{t}&=\beta_{\bol r,\bol r_a}(\mu_0^{1/2}\bol H)+\omega_a\bol Q,
 \end{aligned}
\end{equation}
where $\beta_{\bol r,\bol r_a}=\beta\delta_{\bol r,\bol r_a}$, $\delta_{\bol r,\bol r_a}$ is the Kronecker symbol and $\beta$ is the coupling strength between the antenna emitted wave and the magnetic field.

Finally we turn  our attention to the initial conditions. The initial state vector of Eq. \eqref{24} is
\begin{equation}\label{28}
\bol\psi(\bol r,0)=\bol\psi_0=\begin{bmatrix}
\epsilon_0^{1/2}\bol E_0\\
\mu_0^{1/2}\bol H_0\\
0\\
0
\end{bmatrix}.
\end{equation}
Inclusion of the antenna coupling Eq. \eqref{27} adds to the initial state $\bol\psi_0$ the term $\bol Q(\bol r,0)=\bol Q_a$. The selection of the initial vacuum electromagnetic filed profiles is dictated by the satisfaction of the divergence set of Maxwell equations.
\begin{equation}\label{29}
\bol\div\bol D_0=\bol\div\bol E_0=0,\quad \bol\div\bol B_0=0.
\end{equation}
In that way, the divergence Maxwell equations are guaranteed to be satisfied for $t>0$ along with  $\bol\div\bol J_{cj}=0$ from the charge continuity equation in the continuum limit.

\subsection{Trotter Product Evolution Approximation}\label{sec:2d}
Application of QLA or any other quantum protocol for simulation of electromagnetic wave propagation in a cold inhomogeneous magnetized plasma requires a decomposition  of the $\hat{D}$ operator in Eq.\eqref{24} into simpler matrices,
\begin{equation}\label{30}
\hat{D}=\hat{D}_{vac}+\sum_{j=i,e}[\hat{D}_{\omega_{pj}}+\hat{D}_{\omega_{cj}}],
\end{equation}
with
\begin{align}
\hat{D}_{vac}&=-\frac{c}{2}(I_{2\times2}+\hat{\sigma}_z)\otimes\hat{\sigma}_y\otimes\bol\curl \label{31} \\
\hat{D}_{\omega_{pi}}&=\frac{1}{2}\hat{\sigma}_y\otimes(I_{2\times2}+\hat{\sigma}_z)\otimes\omega_{pi} \label{32} \\
\hat{D}_{\omega_{pe}}&=\frac{1}{2}(\hat{\sigma}_x\otimes\hat{\sigma}_y+\hat{\sigma}_y\otimes\hat{\sigma}_x)\otimes\omega_{pe} \label{33} \\
\hat{D}_{\omega_{ci}}&=\frac{1}{4}(I_{2\times2}-\hat{\sigma}_z)\otimes(I_{2\times2}+\hat{\sigma}_z)\otimes\omega_{ci}\hat{S}_z \label{34} \\
{D}_{\omega_{ce}}&=\frac{1}{4}(I_{2\times2}-\hat{\sigma}_z)\otimes(I_{2\times2}-\hat{\sigma}_z)\otimes\omega_{ce}\hat{S}_z.\label{35}
\end{align}
For simplicity let us assume that all quantities are only $x$-dependent, rendering our model 1D.
The inclusion of $y$- and $z$-dependence is straightforward, following the usual Alternate Direction Iteration (ADI) Cartesian integration procedure with no extraneous couplings of the respective quantum operators.
Then, the curl operator in Eq.\eqref{31} reads
\begin{equation}\label{36}
\bol\curl=\hat{S}_x\hat{p}_x,\quad\hat{S}_x=\begin{bmatrix}
0&0&0\\
0&0&-i\\
0&i&0
\end{bmatrix},\quad\hat{p}_x=-i\pdv{}{x}.
\end{equation}

Trotterizing the total unitary evolution $e^{-i\delta t\hat{D}}$ whose components are given in Eqs.\eqref{30}-\eqref{35} we obtain
\begin{equation}\label{38}
\bol\psi(\bol r,\delta t)=e^{-i\delta t\hat{D}_{vac}}\prod_{j=i,e}e^{-i\delta t\hat{D}_{\omega_{pj}}}e^{-i\delta t\hat{D}_{\omega_{cj}}}\bol\psi_0+\textit{O}(\delta t^2).
\end{equation}
Each of the exponential operators in Eq.\eqref{38} can be written as a product of unitary operators based on the their tensor-fold Pauli structure. Specifically, we have the following diagonalization relations for  the $\hat\sigma_y, \hat\sigma_x, \hat S_x, \hat S_z$ matrices 
\begin{equation}\label{39}
\begin{alignedat}{2}
&\hat\sigma_x=\hat H\hat\sigma_z\hat H,\quad &&\hat\sigma_y=\hat H_y\hat\sigma_z\hat H_y,\\
&\hat S_x=\hat H^{(x)}_y\hat\sigma^{(x)}_z\hat H^{(x)}_y,\quad &&\hat S_z=\hat H^{(z)}_y\hat\sigma^{(z)}_z\hat H^{(z)}_y,
\end{alignedat}
\end{equation}
where $\hat{H}$ is the unitary Hadamard gate, $\hat H_y$ is the unitary variant of Hadamard gate that diagonalizes $\hat\sigma_y$ whereas the unitary set of matrices $\hat H^{(x)}_y, \hat H^{(z)}_y$ and  Hermitian $\hat\sigma^{(x)}_z, \hat\sigma^{(z)}_z$ are the three-dimensional extensions of $\hat H_y$ and $\hat\sigma_z$ for $x$ and $z$ axes respectively,
\begin{widetext}
\begin{equation}\label{43}
\begin{alignedat}{3}
&\hat H=\frac{1}{\sqrt{2}}\begin{bmatrix}
1&1\\
1&-1
\end{bmatrix},\quad &&\hat{H}_y=\frac{1}{\sqrt{2}}\begin{bmatrix}
1&-i\\
i&-1
\end{bmatrix},\quad &&\hat H^{(x)}_y=\frac{1}{\sqrt{2}}\begin{bmatrix}
1&0&0\\
0&1&-i\\
0&i&-1
\end{bmatrix},\\
&H^{(z)}_y=\frac{1}{\sqrt{2}}\begin{bmatrix}
1&-i&0\\
i&-1&0\\
0&0&1
\end{bmatrix},\quad &&\hat\sigma^{(x)}_z=\begin{bmatrix}
0&0&0\\
0&1&0\\
0&0&-1
\end{bmatrix},\quad &&\hat\sigma^{(z)}_z=\begin{bmatrix}
1&0&0\\
0&-1&0\\
0&0&0
\end{bmatrix}.
\end{alignedat}
\end{equation}
\end{widetext}
This enables us to express the unitary exponential of operators \eqref{31}-\eqref{35} using the identities:
\begin{equation}\label{40}
e^{-i\delta t\hat{V_1}\hat{A}\hat{V_1}^\dagger\otimes\hat{V_2}\hat{B}\hat{V_2}^\dagger}=(\hat{V}_1\otimes\hat{V}_2)e^{-i\delta t\hat{A}\otimes\hat{B}}(\hat{V}^\dagger_1\otimes\hat{V}^\dagger_2),
\end{equation}
\begin{equation}\label{41}
e^{-i \delta t I_{2\times2}\otimes\hat{A}}=I_{2\times2}\otimes e^{-i \delta t\hat{A}},
\end{equation}
\begin{equation}\label{42}
e^{-i\frac{\theta}{2}\hat\sigma_i\otimes\hat A}=I_{2\times2}\otimes\cos{(\hat{A}\theta/2)}-i\hat\sigma_i\sin{(\hat{A}\theta/2)}.
\end{equation}

Therefore, the exponential operator $e^{-i\delta t\hat D_{vac}}$ can be written
\begin{equation}\label{44}
e^{-i\delta t\hat D_{vac}}=\hat{C}_{vac}\hat{S}\hat{C}_{vac}
\end{equation}
where the unitary collision operator $\hat{C}_{vac}$ has the form
\begin{equation}\label{45}
\hat{C}_{vac}=I_{2\times2}\otimes\hat H_y\otimes\hat H^{(x)}_y,
\end{equation}
and the advection operator in $x$-direction:
\begin{equation}\label{46}
\hat S=\exp{i(I_{2\times2}+\hat\sigma_z)\otimes\hat\sigma_z\otimes\hat\sigma^{(x)}_zc\delta t\hat{p}_x/2}.
\end{equation}

Similarly, we express the rest of the operators in the Trotterized evolution Eq.\eqref{38} as follows
\begin{equation}\label{47}
e^{-i\delta t\hat D_{\omega_{pi}}}=\hat C_{\omega_{pi}}(\hat{\mathcal{R}}^{(pi)}_z\otimes I_{3\times3})\hat C_{\omega_{pi}},
\end{equation}
where $\theta_{pi}=\omega_{pi}\delta t$, $\hat{C}_{\omega_{pi}}$ is the collision operator
\begin{equation}\label{48}
\hat{C}_{\omega_{pi}}=\hat H_y\otimes I_{2\times2}\otimes I_{3\times3}
\end{equation}
and the $\hat{\mathcal{R}}^{(pi)}_z$ operator is defined through identity \eqref{42} which in principle represents a functional  $\hat{R}_i(\cdot)$ rotations,
\begin{equation}\label{49}
\hat{\mathcal{R}}^{(pi)}_z=[\hat{R}_z(\theta_{pi})\otimes I_{2\times2}]\hat{R}_z(\hat\sigma_z\theta_{pi}).
\end{equation}
For $e^{-i\delta t\hat D_{\omega_{pe}}}$ we obtain
\begin{equation}\label{50}
e^{-i\delta t\hat D_{\omega_{pe}}}=\hat{C}^{(1)}_{\omega_{pe}}(\hat{\mathcal{R}}_z^{(pe)}\otimes I_{3\times3})\hat{C}^{(1)}_{\omega_{pe}}\hat{C}^{(2)}_{\omega_{pe}}(\hat{\mathcal{R}}_z^{(pe)}\otimes I_{3\times3})\hat{C}^{(2)}_{\omega_{pe}}
\end{equation}
with
\begin{align}
\hat{C}^{(1)}_{\omega_{pe}}&=\hat H\otimes\hat H_y\otimes I_{3\times3}, \label{51}\\
\hat{C}^{(2)}_{\omega_{pe}}&=\hat H_y\otimes\hat H\otimes I_{3\times3}, \label{52}\\
\hat{\mathcal{R}}_z^{(pe)}&=\hat{R}_z(\hat\sigma_z\theta_{pe}). \label{53}
\end{align}
We now move to the terms containing the cyclotron angle $\theta_{cj}$,
\begin{equation}\label{54}
\begin{aligned}
e^{-i\delta t\hat D_{\omega_{ci}}}&=\hat{C}_{\omega_{ci}}[I_{4\times4}\otimes\hat{R}_z^{(z)}(\theta_{ci}/2)][I_{2\times2}\otimes\hat{R}_z(\hat\sigma^{(z)}_z\theta_{ci}/2)]\\
&\times\hat{\mathcal{R}}^{(1),(ci)\dagger}_z\hat{\mathcal{R}}^{(2),(ci)\dagger}_z\hat{C}_{\omega_{ci}},
\end{aligned}
\end{equation}
with
\begin{equation}\label{55}
\hat{C}_{\omega_{ci}}=I_{2\times2}\otimes I_{2\times2}\otimes\hat{H}^{(z)}_y
\end{equation}
and operators $\hat{R}_z^{(z)}(\theta_{ci}/2), \hat{\mathcal{R}}^{(1),(ci)}_z,\hat{\mathcal{R}}^{(2),(ci)}_z$ representing $z$-rotation based on the $3\times3$ $\hat\sigma^{(z)}_z$ matrix and functional $z$-rotations respectively,
\begin{align}
\hat{R}_z^{(z)}(\theta_{ci}/2)&=e^{-i\frac{\theta_{ci}}{4}\hat\sigma_z^{(z)}}, \label{56}\\
\hat{\mathcal{R}}^{(1),(ci)\dagger}_z&=e^{i\frac{\theta_{ci}}{4}\hat\sigma_z\otimes I_{2\times2}\otimes\hat\sigma_z^{(z)}}, \label{57}\\
\hat{\mathcal{R}}^{(2),(ci)\dagger}_z&=e^{i\frac{\theta_{ci}}{4}\hat\sigma_z\otimes\hat\sigma_z\otimes\hat\sigma_z^{(z)}}.\label{58}
\end{align}
Finally,
\begin{equation}\label{59}
\begin{aligned}
e^{-i\delta t\hat D_{\omega_{ce}}}&=\hat{C}_{\omega_{ce}}[I_{4\times4}\otimes\hat{R}_z^{(z)}(\theta_{ce}/2)][I_{2\times2}\otimes\hat{R}^\dagger_z(\hat\sigma^{(z)}_z\theta_{ce}/2)]\\
&\times\hat{\mathcal{R}}^{(1),(ce)\dagger}_z\hat{\mathcal{R}}^{(2),(ce)}_z\hat{C}_{\omega_{ci}}.
\end{aligned}
\end{equation}

It is important to note that after we have made the somewhat standard leading-order Trotterized approximation to the total unitary evolution
operator in Eq.\eqref{24}, the evaluations of all the operators in
Eqs.\eqref{44}-\eqref{59} are exact and no further approximations have been made.

Consequently, the fully unitary evolution sequence reads
\begin{widetext}
\begin{equation}\label{60}
\begin{aligned}
&\bol\psi(\bol r,\delta t)=\hat{C}_{vac}\hat{S}\hat{C}_{vac}\hat C_{\omega_{pi}}(\hat{\mathcal{R}}^{(pi)}_z\otimes I_{3\times3})\hat C_{\omega_{pi}}\hat{C}^{(1)}_{\omega_{pe}}(\hat{\mathcal{R}}_z^{(pe)}\otimes I_{3\times3})\hat{C}^{(1)}_{\omega_{pe}}\hat{C}^{(2)}_{\omega_{pe}}(\hat{\mathcal{R}}_z^{(pe)}\otimes I_{3\times3})\hat{C}^{(2)}_{\omega_{pe}}\hat{C}_{\omega_{ci}}[I_{4\times4}\otimes\hat{R}_z^{(z)}(\theta_{ci}/2)]\\
&\times[I_{2\times2}\otimes\hat{R}_z(\hat\sigma^{(z)}_z\theta_{ci}/2)]\hat{\mathcal{R}}^{(1),(ci)\dagger}_z\hat{\mathcal{R}}^{(2),(ci)\dagger}_z\hat{C}_{\omega_{ci}}\hat{C}_{\omega_{ce}}[I_{4\times4}\otimes\hat{R}_z^{(z)}(\theta_{ce}/2)][I_{2\times2}\otimes\hat{R}^\dagger_z(\hat\sigma^{(z)}_z\theta_{ce}/2)]\hat{\mathcal{R}}^{(1),(ce)\dagger}_z\hat{\mathcal{R}}^{(2),(ce)}_z\hat{C}_{\omega_{ci}}\bol\psi_0.
\end{aligned}
\end{equation}
\end{widetext}

\subsection{Quantum encoding and complexity analysis}\label{sec:2e}
Implementation of the Trotterized unitary product formula Eq.\eqref{60} in a digital quantum computer requires spatial discretization. We pursue a qubit lattice algorithm (QLA)  discretization where the evolution \eqref{60} transcends into an interleaved sequence of non-commuting QLA collision $\hat{\mathcal{C}}$  and streaming $\hat{\mathcal{S}}$ operators that recover the Schrodinger-Maxwell equation \eqref{24} to a second order diffusion scheme, $\delta t\sim\delta^2,\,\,\delta x\sim\delta$.  The advantage of this description stems from treating the advection operator $\hat{S}$ in Eq.\eqref{46}, through the QLA streaming operators $\hat{\mathcal{S}}$'s, enabling an efficient quantum implementation\cite{Gourdeau,Succi,Yepez,Koukoutsis}. The rest of the participating operators in Eq.\eqref{60} will comprise the QLA collision operators $\hat{\mathcal{C}}$.

Ultimately, to implement the QLA evolution derived from Eq.\eqref{60} onto a quantum computer we must express the participating operators into elementary quantum gates acting on a set of qubits. We will use two qubit registers. The first encodes the amplitude dimensionality of the state vector $\bol{\psi}$ in Eq.\eqref{24}, hence containing $n_i=4$ qubits with $\{\ket{i}\}$ basis. The second register labels the spatial  discretization. For a one-dimensional lattice with $N$ nodes and a discretization step $\delta$, we will need $n_p=\log_2N$ qubits with basis $\{\ket{p}\}$. Therefore, a total number of $n_{total}=n_p+4$ qubits are required for the complete description of the state $\bol{\psi}$.

Then, the qubit encoding of the state vector $\bol{\psi}$ reads,
\begin{equation}\label{12}
\ket{\bol\psi}=\sum_{p=0}^{N-1}\sum_{i=0}^{11}\psi_{0ip}\ket{i}\ket{p},
\end{equation}
with amplitudes $\psi_{ip}$ characterize the $i$-component of the state vector $\bol{\psi}$ in the lattice site $p$. The quantum state $\ket{\bol\psi}$ is normalized to the square root of the initial (constant) electromagnetic energy so that $\sum_{i,p}\abs{\psi_{0ip}}^2=1$.

Establishing the required circuit width (total number of qubits) for the quantum encoding of our state, we proceed to analyze the decomposition scaling (circuit depth) of operators in Eq.\eqref{60} into simple one-qubit and CNOT gates to $n_{total}$. All the unitary collision $\hat{C}$'s operators are in tensor product of elementary single-qubit gates like the Hadamard gate $\hat{H}$ and rotation gate $\hat{H}_y=\hat\sigma_z\hat{R}_x(\pi/2)$ whereas the $\hat{H}_y^{(z)},\hat{H}_y^{(x)}$  two-level gates can be easily implemented within simple, one-qubit gates. In addition, those operators act solely in the 4-qubit amplitude register $\{\ket{i}\}$, resulting to constant scaling and can be implemented in the worst case scenario as $\textit{O}(k\cdot4^2),\,\,k\in\mathcal{N}$. The integer $k$ accounts for the total number of collision operators $\hat{C}$ in Eq.\eqref{60}. As far as the unitary rotation operators which contain the plasma inhomogeneity are concerned, they are all diagonal and can be decomposed into simpler two-level $z$-rotations or directly implemented within $\textit{O}(m\cdot2^{n_{total}+1})$ CNOTs and single-qubit gates\cite{Bullock}. As previous, the natural number $m$ now accounts for the total number of those diagonal inhomogeneous operators in Eq.\eqref{60}. Finally, the QLA streaming  $\hat{\mathcal{S}}$ operators offer the advantage of implementing  the associated advective operator $\hat{S}$ as a quantum walk\cite{Succi}. The explicit circuit implementation of this quantum walk into a quantum computer is presented in Refs.[\onlinecite{Koukoutsis}], [\onlinecite{Gourdeau}]. The QLA streaming operators act only in the spatial discretization register $\{\ket{p}\}$, controlled by the $\{\ket{i}\}$ qubits, so based on the results of Refs.[\onlinecite{Koukoutsis}] and  [\onlinecite{Gourdeau}] they are expected to scale as  $\textit{O}(l\cdot n^2_{p}),\,\,l\in\mathcal{N}$.

Consequently, the total quantum implementation scaling of the QLA discretization scheme of unitary evolution \eqref{60} is expected to be $\textit{O}(32m\cdot2^{n_p}+l\cdot n^2_{p}+16k)$. For fusion relevant applications the inhomogeneity plasma profile is localized, enabling us to reduce the encoding cost of the inhomogeneous diagonal rotation operators to $\textit{O}[poly(n_p)]$ which in turn implies that the total implementation cost of our algorithm scales polynomially $\textit{O}[(poly(n_p)]$ with the number of qubits in the $p$-register. This polynomial scaling promotes QLA as prominent candidate for implementation in real quantum hardware in the near future. 

\subsection{Discussion}\label{sec:2f}
Comparing the Schrodinger representation of Maxwell equations for inhomogeneous non-dispersive media Eq.\eqref{1} with Eq.\eqref{24} for the magnetized plasma, it seems that the latter supports more complexity due to the dimensionality of the state vector $\bol\psi$. But, in contrast with the optical case where the respective spatial displacement operator interferes with the inhomogeneity of the refractive index (see Eq.\eqref{2}) the respective exponential of operator $\hat{D}_{vac}$ in Eq.\eqref{31} is explicitly decomposed without implicit dependence on the inhomogeneity plasma profile which is reflected in the plasma frequencies. As a consequence, the expected QLA  will be free of the non-unitary potential operators $\hat{V}$ such those introduced in Refs.[\onlinecite{Vahala3,Vahala4,Vahala1}], resulting in a fully unitary product sequence similar to that of a homogeneous medium\cite{Vahala2}.

Subsequently, a vacuum QLA sequence denoted as $\hat{U}^{vac}_X$ can be immediately employed to calculate the term $e^{-i\delta t\hat{D}_{vac}}$ in the Trotterized evolution approximation of $e^{-i\delta t\hat{D}}$,
\begin{equation}\label{37}
\begin{aligned}
e^{-i\delta t\hat{D}}&=e^{-i\delta t\hat{D}_{disp}}e^{-i\delta t\hat{D}_{vac}}+\textit{O}(\delta t^2)\\
&=e^{-i\delta t\hat{D}_{disp}}\hat{U}_{X}^{vac}+\textit{O}(\delta t^2).
\end{aligned}
\end{equation}
Implementation of the dispersive part $e^{-i\delta t\hat{D}_{disp}}$, where $\hat{D}_{disp}=\sum_{j=i,e}\hat{D}_{\omega_{pj}}+\hat{D}_{\omega_{cj}}$ can be performed in parallel with the QLA. The main advantage of this approximation is that we can decide whether to classically compute the $\hat{U}_{X}^{vac}\bol \psi_0$, store the information and proceed with a follow up quantum computation for the $e^{-i\delta t\hat{D}_{disp}}$ term resulting in a hybrid computation, or purely quantum computing the whole sequence based on the quantum encoding of QLA as described in Sec.\ref{sec:2e}.

In addition, the unitary QLA derived from evolution sequence \eqref{60} conserve the extended electromagnetic energy Eq.\eqref{25} and the divergence conditions. Thus, our full-wave scheme can be extended beyond the usual plane-wave or monochromatic wave approximations. This is very important in the case of fusion plasmas where the RF waves that are used for plasma heating and current drive are wave-packets that are localized in space and of finite duration in time. The interaction of the inhomogeneity plasma profile with the envelope of the carrier wave, as well as with the individual components that a spatially confined beam consists of, will lead to complex electromagnetic structures that will affect the current densities in the dispersive plasma. More importantly, those transport effects correspond to energy transfer from the initial electromagnetic fields to the current density fields and  can be explicitly measured due to Eq.\eqref{25} which describes the extended electromagnetic energy. Hence, examination of wave packet propagation in plasmas is relevant to realistic fusion experiments. For instance, an initial X-wave polarization $\bol E_0=E_y(k_x x)\hat{\bol y}$ profile, the scattering from a two dimensional $x-y$ plasma inhomogeneity will generate the electromagnetic  fields  $\bol E=E_x(k_x x,k_y y, \omega_X t)\hat{\bol x}+E_y(k_x x,k_y y, \omega_X t)\hat{\bol y}$ and $\bol B=B_z(k_x x,k_y y, \omega_X t)\hat{\bol z}$ but most importantly will produce the conductivity current density $\bol J_{cj}=J_{xcj}(k_x x,k_y y, \omega_X t)\hat{\bol x}+J_{ycj}(k_x x,k_y y, \omega_X t)\hat{\bol y}$ to satisfy $\div \bol E=\div \bol B=\div \bol J_{cj}=0$.

Given the fact that the QLA scales linearly with the number of processors and its quantum variant is probably expected to scale as $\textit{O}[poly(n_p)]$, it is evident that our considerations pose  a strong alternative to the cost-inefficient FDTD methods, particularly in 2D and 3D.

On the other hand, it may be necessary to further manipulate the evolution sequence \eqref{60} for an optimized QLA to be produced.\cite{Yepez2,unpublished} Therefore, considerable research is required before applying the QLA for simulation of wave propagation into a plasma characterized by fusion-reactor parameters. 

\section{Example: QLA for scattering from 2D scalar non-dispersive dielectric objects}\label{sec:3}
Although the analytical and algorithmic establishments in Sec.\ref{sec:2} should result in an efficient quantum computer code for electromagnetic wave propagation in cold inhomogeneous magnetized plasmas, much work remains to be done in optimizing the qubit presentation of a QLA code for Eq.\eqref{60} before tackling the propagation of such fusion relevant RF wave-packets in plasma.

It is thus instructive to first investigate our Maxwell QLA code capabilities and behavior for the scattering of an electromagnetic pulse from a non-dispersive 2D inhomogeneous dielectric object, and we shall observe some interesting physics arising from the initial value simulations.

\subsection{The algorithm}\label{sec:3a}
\begin{figure}[htbp]

\subfloat[\label{subfig:1a}]{%
  \includegraphics[width=1.0\columnwidth]{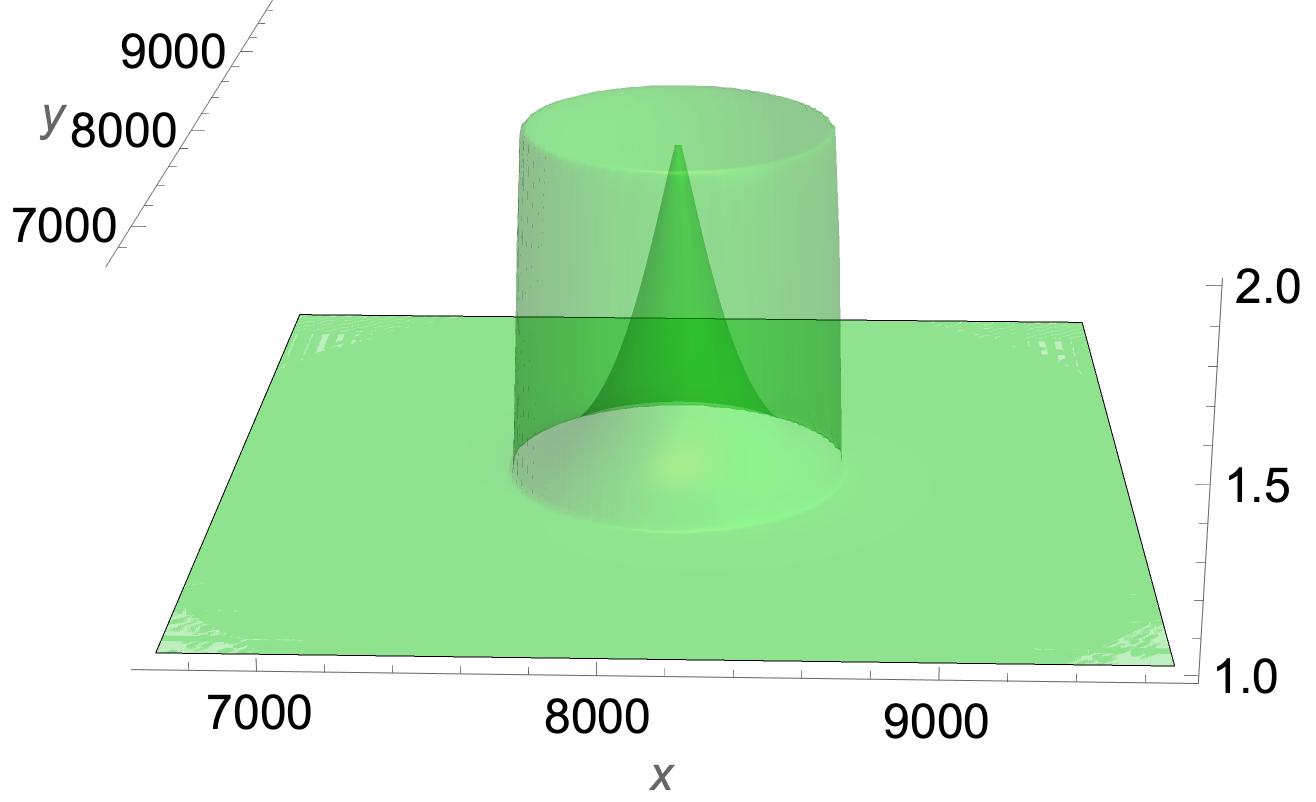}%
}\hfill
\subfloat[\label{subfig:1b}]{%
  \includegraphics[width=1.0\columnwidth]{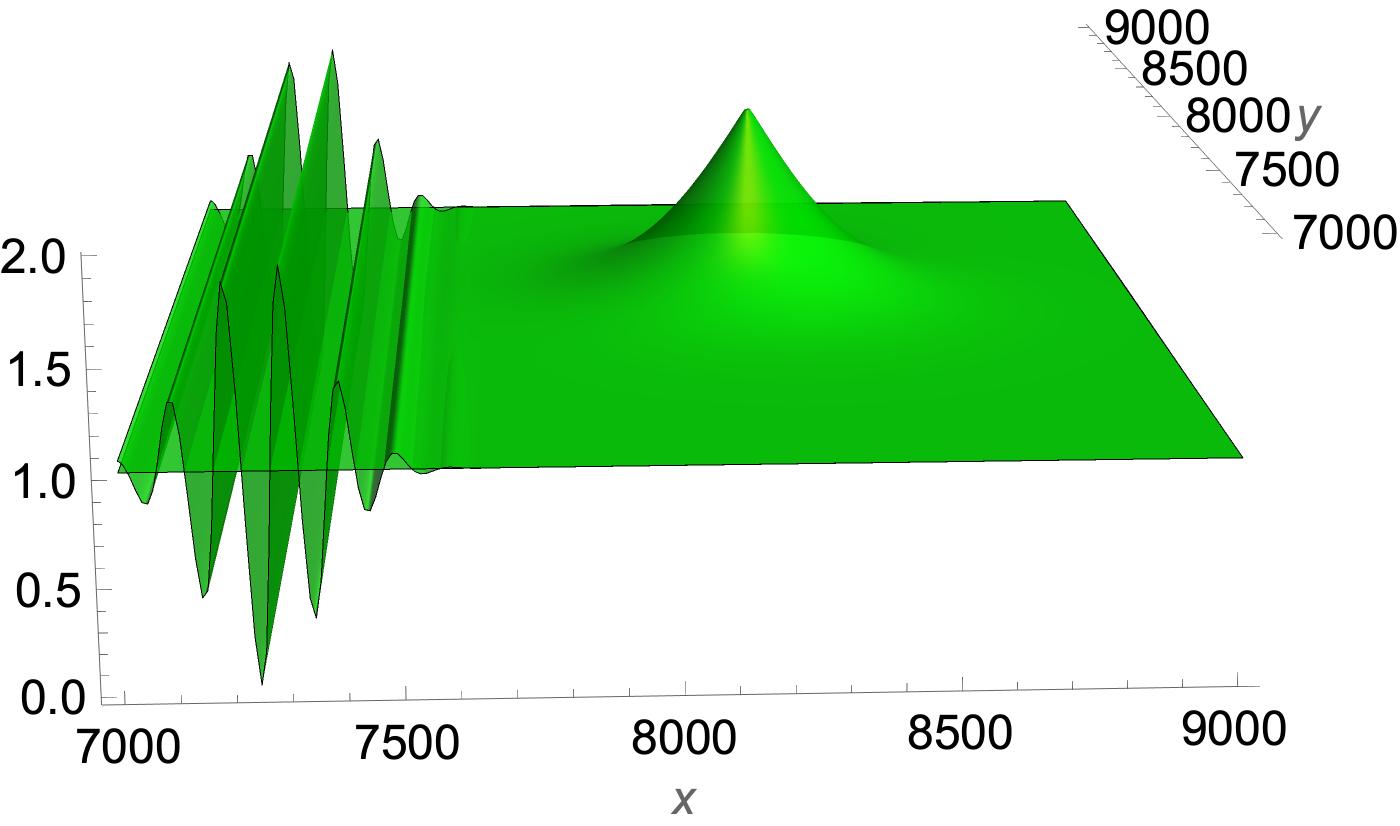}%
}
\caption{Two different inhomogeneity refractive index profiles $1 \le n(x,y) \le 2$ and the electric field $E_{z0}(x)$ of the incident wave-packet. The cylinder dielectric has strong spatial gradient near the vacuum-dielectric interface, while the conic dielectric has very weak spatial gradients.  In Fig.\ref{subfig:1a} these two profiles are shown superimposed.  In Fig.\ref{subfig:1b} the conic dielectric is shown together with the incident wave-packet (arbitrary normalization).
}\label{fig:1}
\end{figure}

To showcase what a QLA sequence looks like and what we expect to obtain from the ``QLAzation" of Eq.\eqref{60}, we briefly present the algorithmic scheme for a 2D $x-y$ scattering of a wave-packet from a scalar but non-dispersive localized inhomogeneities with refractive index $n = n(x,y)$, as displayed in Fig.\ref{fig:1}. The shape of the inhomogeneities, can be related to cylindrical filaments or smooth localized concentrations of plasma density.

In our reduced case of non-dispersive dielectric, QLA is a discrete representation of unitary representation of Maxwell equations \eqref{1} which, at a mesoscopic level, uses an appropriately chosen interleaved sequence of three non-commuting operators. Two of the operators are unitary collision and streaming operators – the collision operator entangles the on-site qubits and the streaming operator propagates the entangled state through the lattice. The gradients in the medium constitutive properties are included via a third operator referred to as a potential operator.

For 2D $x-y$ scattering of electromagnetic fields for a scalar dielectric state vector that evolves unitarily is 
\begin{equation}\label{4}
\bol{q}=\begin{bmatrix}
nE_x\\
nE_y\\
nE_z\\
\mu_0^{1/2}H_x \\
\mu_0^{1/2}H_y \\
\mu_0^{1/2}H_z
\end{bmatrix}=\begin{bmatrix}
q_0\\
q_1\\
q_2\\
q_3\\
q_4\\
q_5
\end{bmatrix}.
\end{equation}
In (diagonal) tensor dielectric media one would simply have $q_0 \rightarrow n_xE_x$, $q_1 \rightarrow n_yE_y$, $q_2 \rightarrow n_zE_z$.

The decomposition of the electromagnetic Schrodinger equation \eqref{1} in Cartesian components is
\begin{equation}\label{5}
\begin{aligned}
&\pdv{q_0}{t}=\frac{1}{n}\pdv{q_5}{y},\quad \pdv{q_1}{t}=\frac{1}{n}\pdv{q_5}{x},\quad \pdv{q_2}{t}=\frac{1}{n}\Big[\pdv{q_4}{x}-\pdv{q_3}{y}\Big],\\
&\pdv{q_3}{t}=\pdv{(q_2/n)}{y},\quad \pdv{q_4}{t}=\pdv{(q_2/n)}{x},\\
&\pdv{q_5}{t}=-\pdv{(q_1/n)}{x}+\pdv{(q_0/n)}{n_y}.
\end{aligned}
\end{equation}
For the discrete QLA, using the Alternating Directions Implicit (ADI) integration, the unitary collision operators in the x and y directions are
\begin{equation}\label{6}
\hat{C}_X=\begin{bmatrix}
 1&0&0&0&0&0\\
 0&\cos{\theta_0}&0&0&0&-\sin{\theta_0}\\
 0&0&\cos{\theta_0}&0&-\sin{\theta_0}&0\\
 0&0&0&1&0&0\\
 0&0&\sin{\theta_0}&0&\cos{\theta_0}&0\\
 0&\sin{\theta_0}&0&0&0&\cos{\theta_0}
\end{bmatrix},
\end{equation}
\\
\begin{equation}\label{7}
\hat{C}_Y=\begin{bmatrix}
 \cos{\theta_0}&0&0&0&0&\sin{\theta_0}\\
 0&1&0&0&0&0\\
 0&0&\cos{\theta_0}&\sin{\theta_0}&0&0\\
 0&0&-\sin{\theta_0}&\cos{\theta_0}&0&0\\
 0&0&0&0&1&0\\
 -\sin{\theta_0}&0&0&0&0&\cos{\theta_0}
\end{bmatrix}.
\end{equation}
with collision angle $\theta_0 = \delta /4n$.
The form of $\hat{C}_X$ can be readily discerned from the coupling of the $\pdv{}{t}$ with $\pdv{}{x}$ derivatives in \eqref{5}:  $q_1 - q_5$, and
$q_2 - q_4$, as well as the respective collision angle.  Similarly for the unitary matrix $\hat{C}_Y$. 

We now define the unitary streaming operator $\hat{S}_{ij}$ which shifts the amplitudes $\{q_i, q_j\}$ one lattice unit, either in the $x$ or in the {y}  direction, 
while leaving all the other amplitudes unaffected. Then the collide-stream sequence along each direction is,

\begin{widetext}
\begin{equation}
\begin{aligned}\label{8}
\hat{U}_X&=\hat{S}^{+x}_{25}\hat{C}^\dagger_X\hat{S}^{-x}_{25}\hat{C}_X\hat{S}^{-x}_{14}\hat{C}^\dagger_X\hat{S}^{+x}_{14}\hat{C}_X\hat{S}^{-x}_{25}\hat{C}_X\hat{S}^{+x}_{25}\hat{C}^\dagger_X\hat{S}^{+x}_{14}\hat{C}_X\hat{S}^{-x}_{14}\hat{C}^\dagger_X\\
\hat{U}_Y&=\hat{S}^{+y}_{25}\hat{C}^\dagger_Y\hat{S}^{-y}_{25}\hat{C}_Y\hat{S}^{-y}_{03}\hat{C}^\dagger_Y\hat{S}^{+y}_{03}\hat{C}_Y\hat{S}^{-y}_{25}\hat{C}_Y\hat{S}^{+y}_{25}\hat{C}^\dagger_Y\hat{S}^{+y}_{03}\hat{C}_Y\hat{S}^{-y}_{03}\hat{C}^\dagger_Y.
\end{aligned}
\end{equation}
\end{widetext}
It should be noted that the first set of four collide-stream operators in $\hat{U}_X$ and $\hat{U}_Y$ would yield \eqref{5} to first order in $\delta$. An in-depth analysis on derivation of the QLA sequences Eq.\eqref{8} can be found in Refs.[\onlinecite{Vahala1,Vahala3,Vahala4,Yepez2,unpublished}] and in references therein.

The terms in \eqref{5}, containing the derivatives of the refractive index, are recovered through the following potential operators
\begin{equation}\label{9}
\hat{V}_X=\begin{bmatrix}
1&0&0&0&0&0\\
0&1&0&0&0&0\\
0&0&1&0&0&0\\
0&0&0&1&0&0\\
0&0&-\sin{\beta_0}&0&\cos{\beta_0}&0\\
0&\sin{\beta_0}&0&0&0&\cos{\beta_0}
\end{bmatrix}
\end{equation}
and
\begin{equation}\label{10}
\hat{V}_Y=\begin{bmatrix}
1&0&0&0&0&0\\
0&1&0&0&0&0\\
0&0&1&0&0&0\\
0&0&\cos{\beta_1}&\sin{\beta_1}&0&0\\
0&0&0&0&1&0\\
-\sin{\beta_1}&0&0&0&0&\cos{\beta_1}
\end{bmatrix}.
\end{equation}
The  angles $\theta_0=\delta /4n$, $\beta_0=\delta^2\frac{\partial n/\partial x}{n^2}$, and  $\beta_1=\delta^2\frac{\partial n/\partial y}{n^2}$, that appearing in matrices \eqref{6}, \eqref{7}, \eqref{9}, and \eqref{10} are chosen so that the discretized system reproduces \eqref{5} to order $\textit{O}(\delta^2)$.

The evolution of the state vector $\bol{q}$ from time $t$ to $t+\Delta{t}$ is given by,
\begin{equation}\label{11}
\bol{q}(t+\Delta{t})=\hat{V}_Y\hat{V}_X\hat{U}_Y\hat{U}_X\bol{q}(t).
\end{equation}
Note that the external potential operators $\hat{V}_X, \hat{V}_Y$, as given above, are not unitary. Quantum implementation of the non-unitary potential operators $\hat{V}_X,\hat{V}_Y$ can be handled using the Linear Combination of Unitaries (LCU) method.\cite{Childs} We direct the reader to Ref. [\onlinecite{Koukoutsis}] for a detailed discussion on the quantum circuit implementation of these QLA non-unitary operators.

A detailed analysis of the QLA  for the more general case of a bi-axial medium along with simulation results for scattering of Gaussian pulses can be found in Ref. [\onlinecite{unpublished}].

\subsection{QLA simulation results}\label{sec:3b}
In all simulations, the total energy is conserved to the seventh significant digit.
A numerical study of errors with respect to spatial resolution was performed in Ref.[\onlinecite{Yepez}]. It indeed verified that the QLA performs better than $2$nd order accuracy. This scaling was further verified in Ref.[\onlinecite{Taylor}] for spinor BECs. In addition, from current discrete simulation 2D QLA runs\cite{Vahala2,unpublished}, it appears that divergence cleaning is not required as QLA divergence errors are spatially localized and do not accumulate. We also reiterate that in applications of QLA to nonlinear spinor Bose-Einstein condensates, the QLA produced an algorithm that was ideally parallelized to all available cores on a classical supercomputer (over $750,000$ cores on the now-retired IBM Blue Gene/$Mira$ supercomputer at Argonne).
\begin{figure}[htbp]
\subfloat[\label{subfig:2a}]{%
  \includegraphics[width=1.0\columnwidth]{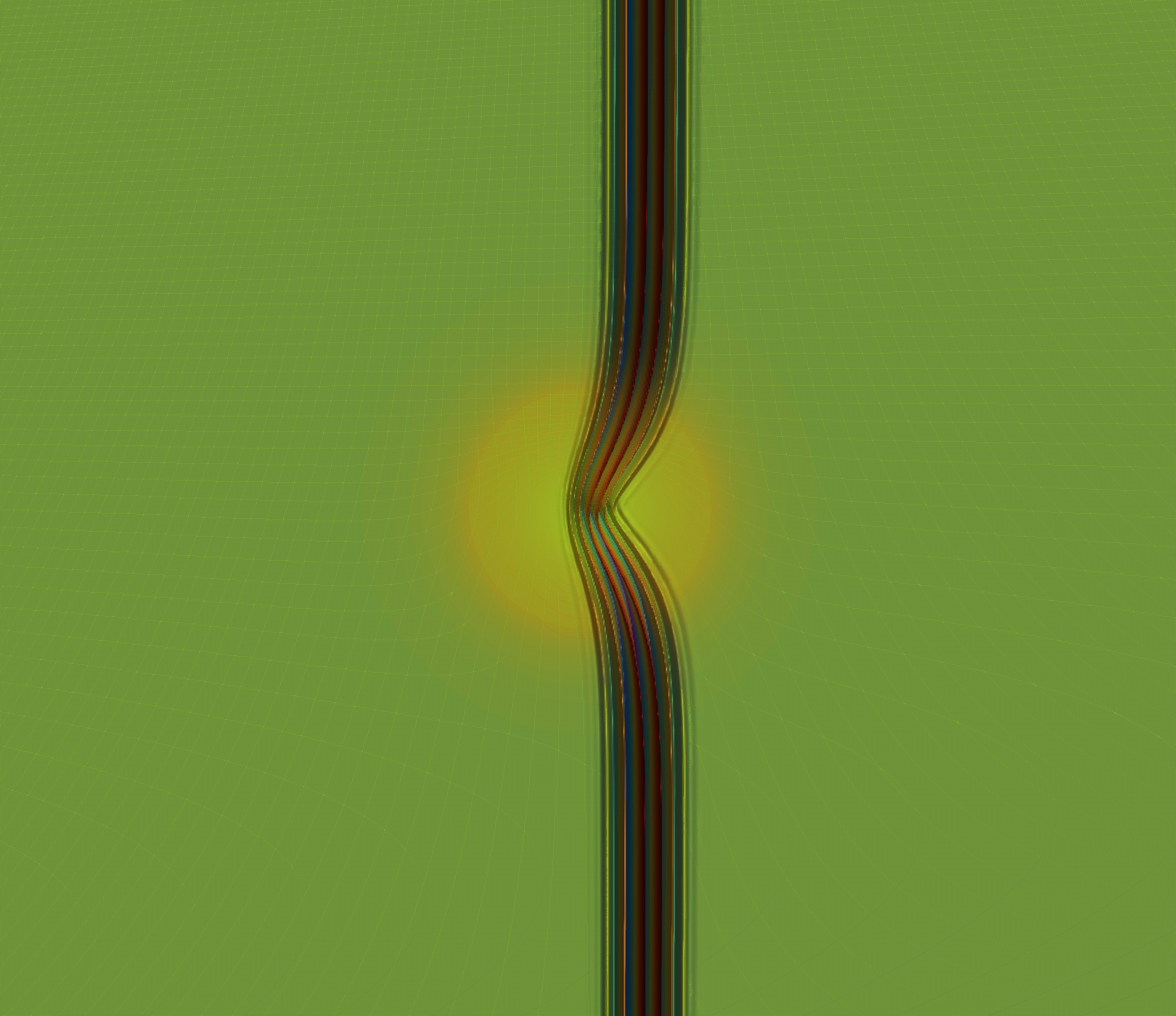}%
}\hfill
\subfloat[\label{subfig:2b}]{%
  \includegraphics[width=1.0\columnwidth]{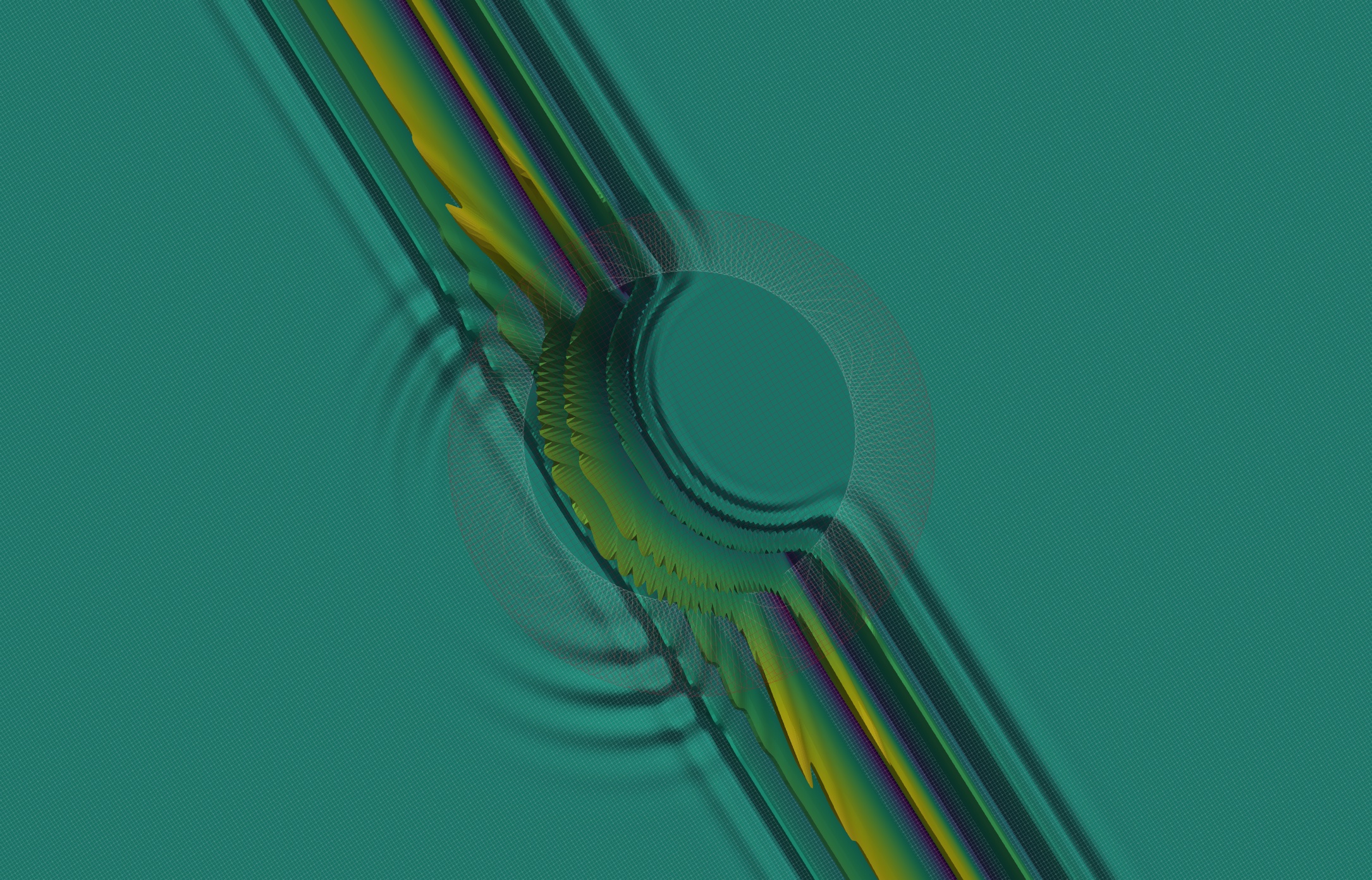}%
}
\caption{QLA scattering simulation of $z$-component of an electromagnetic pulse, $E_{z0}$ off a  dielectric inhomogeneity in the shape of a cone (Fig.\ref{subfig:2a}), versus a cylindrical dielectric (Fig.\ref{subfig:2b}).  The perspective is looking down the z-axis onto the x-y plane.  The full-wave simulation for the wave-cylinder encounter reveals strong initial reflection phenomena whereas the reflection is very weak in the cone case. This differentiation in the wave behavior is directly related to the steepness of the inhomogeneity gradient. The weak reflected wave from the cone corresponds to asymptotic WKB type of solution.}\label{fig:2}
\end{figure}

The initial electromagnetic wave-packet $\bol u_0=(E_{z0}(x), -B_{y0}(x))^T$ is a Gaussian envelope with internal oscillations, Fig.\ref{subfig:1b}.
The wave-packet propagates in the $x$-direction, from a vacuum $n=1$ towards a localized dielectric inhomogeneous object with $n_{max}(x,y)=2$. This polarization satisfies the initial divergence conditions.  As the 1D vacuum wave-packet interacts with the 2D refractive index of the dielectric. the $B_y$ field now becomes 2D, with $B_y(x,y,t)$.  This self-consistently generates a $B_x(x,y,t)$ so that $\nabla \cdot \bol{B} = 0$ as well as a 2D $E_z(x,y,t)$.  Throughout the QLA scattering simulation, $\nabla \cdot \bol{B}$ is monitored and is non-zero in very small isolated spatial regions with some time instants in which 
$max_{x,y} \abs{\nabla \cdot \bol{B} / \bol{B_0}} \leq 0.006$.  $\nabla \cdot \bol{D}$ is identically zero throughout the simulation.
[For initiial $E_{y0}(x)$-polarization, 2D QLA simulations retain $\nabla \cdot \bol{B} = 0$ identically zero for all time.]

In Fig.\ref{fig:2}, the wave-packet has interacted with the dielectric object.   The viewpoint is looking down from the $z-$axis  onto the $x-y$ plane.  The apex of the cone is seen as a white dot, while the interior of the dielectric cylinder is in a somewhat darker color than the surrounding vacuum.   In the case of a dielectric cone, Fig.\ref{subfig:2a}, there is a mild slowing down of that part of the packet that is around the apex of the cone - since the phase velocity is reduced to $c/n(x,y)$.  But more importantly, one does  not see any reflected part of the packet from the slowly varying boundary region between vacuum and dielectric.  Basically the propagation is WKB-like.
On the other hand there are immediate reflection fronts emitted back into the vacuum from the interaction of the wave-packet's oscillation peaks with the steep refractive index gradients in the boundary region of vacuum and cylinder dielectric, Fig.\ref{subfig:2b}.  There is also considerable retardation in the oscillation peaks within the dielectric cylinder as the refractive index away from the boundaries are $n=2$.

\begin{figure}[htbp]
\subfloat[\label{subfig:3a}]{%
  \includegraphics[width=1.0\columnwidth]{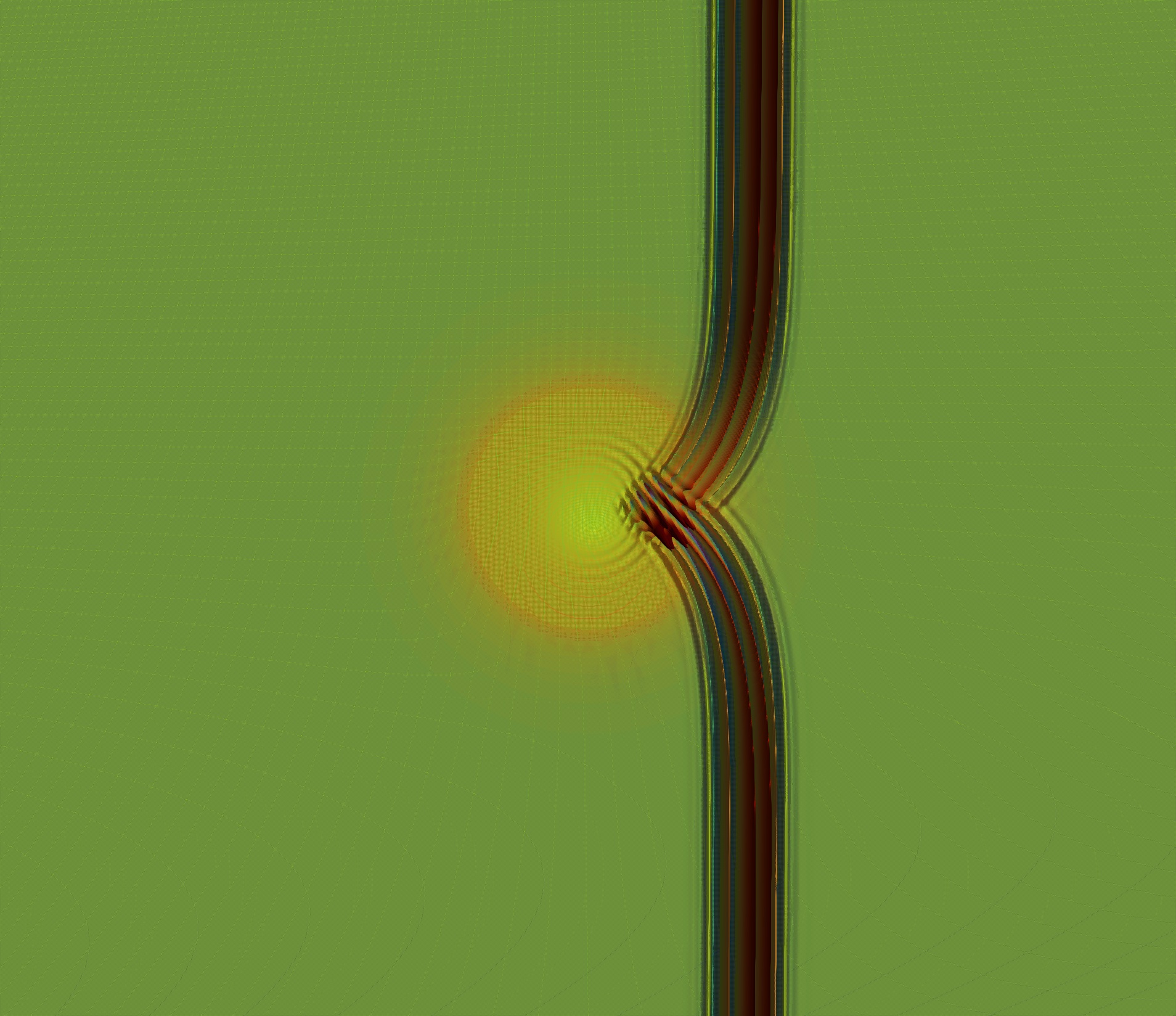}%
}\hfill
\subfloat[\label{subfig:3b}]{%
  \includegraphics[width=1.0\columnwidth]{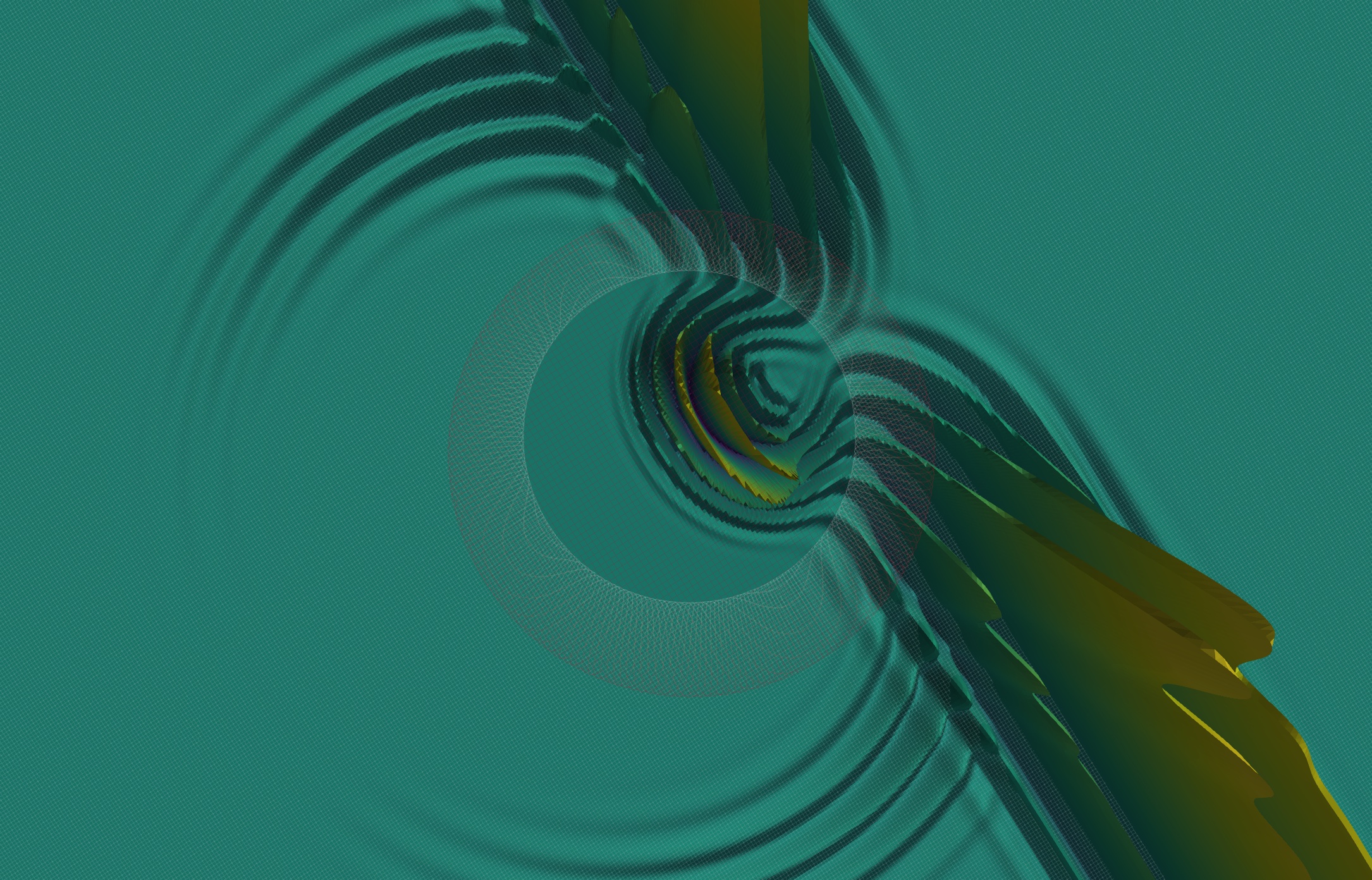}%
}
\caption{The propagation of the transmitted wave within the conical and cylindrical dielectrics. The wave propagation is now distorted because the initial wave crests along the $y-axis$ diffract on the dielectric boundary. In both cases, Figs.\ref{subfig:3a}, \ref{subfig:3b}, transmitted bounded modes are observed towards the exit point to vacuum.}\label{fig:3}
\end{figure}

As mentioned earlier, the transmitted component of the  initial wave-packet propagates into the respective dielectrics with phase velocity
\begin{equation}\label{extra3}
v_{ph}=\frac{c}{n(x,y)}
\end{equation}
because there is no dispersion in the media. However, the wave crests and the envelope along the $y$-direction possess different phase velocities during their propagation in the dielectric resulting in a lag between the interior and outer wave components.Ultimately, both dielectrics exhibit
complex diffraction patterns outside the dielectric as well as bounded eigenmodes within the latter. This behavior is clearly depicted in Fig.\ref{fig:3}.

\begin{figure}[htbp]
\subfloat[\label{subfig:4a}]{%
  \includegraphics[width=1.0\columnwidth]{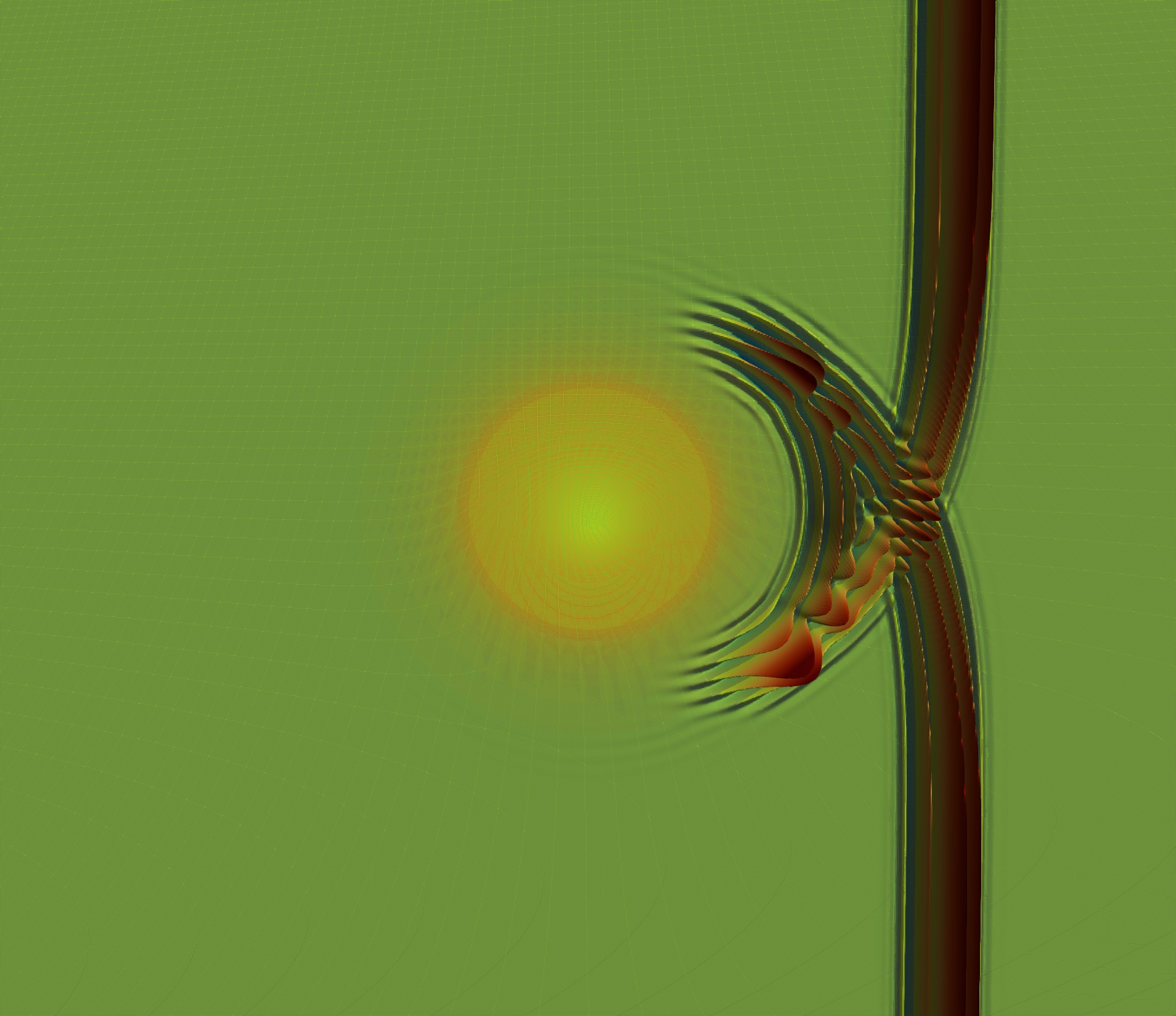}%
}\hfill
\subfloat[\label{subfig:4b}]{%
  \includegraphics[width=1.0\columnwidth]{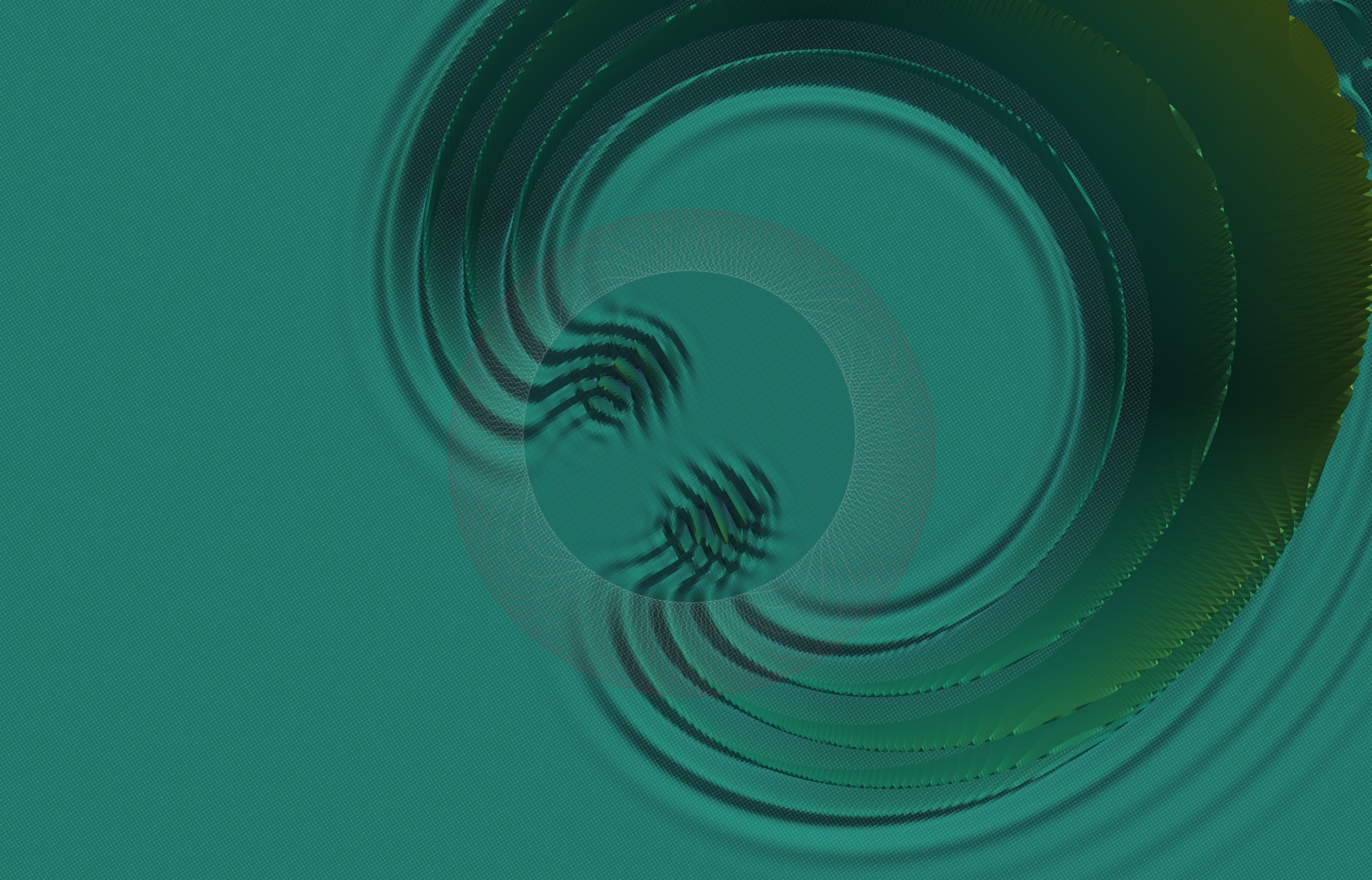}%
}
\caption{The absence of internal reflections from the conical dielectric Fig.\ref{subfig:4a} versus the internal reflections from the cylindrical dielectric Fig.\ref{subfig:4b}. Similar to the behavior of the primary reflections in Fig.\ref{fig:2} the inhomogeneity gradient of the dielectrics plays a pivotal role on the strength of the internal reflection.}\label{fig:4}
\end{figure}

As the bounded modes within the dielectric approach the vacuum boundary, the rapid change in the cylindrical dielectric object produces a secondary internal reflection that propagates back inside the cylinder. For the cone case, the slowly varying transition between the different regions contributes a negligible secondary reflection. Those secondary reflections, along with the secondary propagating wave-fronts in the vacuum region are presented in Fig.\ref{fig:4}.

The back and forth succession from Fig.\ref{fig:4} to Fig.\ref{fig:2}  through  higher order internal reflections in the cylindrical dielectric results in a radiating temporal pattern.  It should be reminded that QLA is an initial value solver giving the temporal (and transient) evolution of the scattered field without the introduction of any internal boundary conditions to handle vacuum-dielectric effects. Even though the simulations are for non-dispersive dielectrics they reveal that the QLA accurately grasps the interconnection of the transient behavior of waves with the inhomogeneity profile. Extending those  considerations to inhomogeneous fusion plasma will provide insights in the temporal evolution of the electromagnetic fields and the species current densities (see the state vector $\bol\psi$ in Eq.\eqref{24}) that potentially could affect the heating efficiency and the energy transfer.

\section{Conclusions}\label{sec:4}
The contributions of this paper are: $(1)$ the analytical formulation of Maxwell equations in a magnetized plasma, Eq.\eqref{24}, as a Schrodinger equation, and $(2)$ a fully unitary QLA representation of this augmented Schrodinger equation indicating a polynomial scaling for implementation in a quantum computer that can be tested on present day classical computers.

The augmented Schrodinger representation has advantages over the standard Helmholtz formulation\cite{Ram2,Ram3} both in the regularity of the spatial derivative of the fields as well as in the construction of formal solutions. The Hermitian structure of the full operator $\hat{D}$ permits a normal mode decomposition of the solution in terms of the eigenfunctions $\bol\phi(\bol r,\lambda)$ of $\hat{D}$ operator with $\lambda$ being the respective eigenvalues. This is very important in cases where the inhomogeneous plasma profile does not possess a simple symmetry. In addition, the unitary evolution of Eq.\eqref{24} explicitly preserves an extended electromagnetic energy integral \eqref{25} beyond the usual Landau and Brillouin approximations\cite{Jackson}.

While various quantum simulation schemes can be devised for the solution of the augmented Schrodinger equation \eqref{24} for wave propagation in a cold magnetized plasma we are currently pursuing a QLA scheme by expressing the energy preserving evolution as the unitary product formula \eqref{60}. This decomposition is deemed suitable for construction of a fully unitary QLA,which no longer requires the introduction of potential operators, and their subsequent quantum encoding. Our findings support that the produced QLA sequence of unitary collision-streaming operators could be implemented on a quantum computer with polynomial scaling in respect with the number of qubits $n_p=\log_2{N}$ required to describe the $N$ lattice cites.

To benchmark the capabilities of QLA we present here the two dimensional scattering of 
a wave-packet from either a cylindrical or a conical scalar, inhomogeneous non-dispersive dielectrics.  For the conic dielectric there are weak spatial gradients in the layer connecting the vacuum to the dielectric.  As a result, there is negligible reflection at the first encounter of the wave packet with the dielectric and then following the interaction with the steep cone apex there is no internal reflections within the dielectric.  This results in a simple scattered field from the cone.  However, for the cylindrical dielectric, the sharp (but continuous) gradient in the layer connecting the dielectric to the vacuum, yields an immediate reflected wave front from the first interaction of the wave packet with the dielectric followed by subsequent reflection/transmission of the wave packet at the dielectric-vacuum layer. This leads to quite complex interference in the scattered fields. 

We are now exploring QLA simulations of the wave propagation in a cold magnetized (dispersive) plasma, exploiting the QLA operator splitting approach. While only the $x$-dependent fully unitary QLA is presented here, the use of the Alternating Direction Implicit (ADI) integration scheme will permit extensions to fully 3D simulations. Moreover, the fact that QLA is ideally parallelized on classical supercomputers together with the polynomial scaling of its quantum implementation yields a pathway for high fidelity simulation results and possibly a hybrid classical-quantum computation model.


%
%

%

\begin{acknowledgments}
This work has been carried out within the framework of the EUROfusion Consortium, funded by the European Union via the Euratom Research and Training Programme (Grant Agreement No 101052200 — EUROfusion). Views and opinions expressed are however those of the authors only and do not necessarily reflect those of the European Union or the European Commission. Neither the European Union nor the European Commission can be held responsible for them.
This research was partially supported by Department of Energy grants DE-SC0021647, DE-FG0291ER-54109, DE-SC0021651, DE-SC0021857, and DE-SC0021653.   This research used resources of the National Energy Research Scientific Computing Center (NERSC), a U.S. Department of Energy Office of Science User Facility located at Lawrence Berkeley National Laboratory, operated under Contract No. DE-AC02-05CH11231 using NERSC award FES-ERCAP0020430.
\end{acknowledgments}

\section*{Author Declarations}

\subsection*{Conflict of Interest}
The authors have no conflicts to disclose.

\subsection*{Author Contributions}
\textbf{Efstratios Koukoutsis}: Conceptualization (lead); Formal analysis (lead); Methodology (equal); Investigation (equal); Writing - original draft (lead); Writing - review $\&$ editing (equal).
\textbf{Kyriakos Hizanidis}: Methodology (equal); Supervision (supporting); Investigation (supporting); Writing - review $\&$ editing (equal); Funding acquisition (equal). 
\textbf{George Vahala}:  Conceptualization - QLA (lead); Methodology (equal); Investigation - QLA (lead); Visualization (equal); Writing - review $\&$ editing (equal); Funding acquisition (equal).
\textbf{Min Soe}: Software - QLA MPI $\&$ Graphics routines (lead); Visualization (equal); Funding acquisition (equal).
\textbf{Linda Vahala}: Data curation - data analysis (lead), Writing - review $\&$ editing (equal); Funding acquisition (equal).
\textbf{Abhay K. Ram}: Methodology (equal); Investigation - physics (equal); Writing - review $\&$ editing (equal); Funding acquisition (equal).

\section*{Data availability}
The data that support the findings of this research are available from the corresponding author upon reasonable request. 

\def\bibsection{\section*{References}} 
\providecommand{\noopsort}[1]{}\providecommand{\singleletter}[1]{#1}%


\begin{thebibliography}{34}%
\makeatletter
\providecommand \@ifxundefined [1]{%
 \@ifx{#1\undefined}
}%
\providecommand \@ifnum [1]{%
 \ifnum #1\expandafter \@firstoftwo
 \else \expandafter \@secondoftwo
 \fi
}%
\providecommand \@ifx [1]{%
 \ifx #1\expandafter \@firstoftwo
 \else \expandafter \@secondoftwo
 \fi
}%
\providecommand \natexlab [1]{#1}%
\providecommand \enquote  [1]{``#1''}%
\providecommand \bibnamefont  [1]{#1}%
\providecommand \bibfnamefont [1]{#1}%
\providecommand \citenamefont [1]{#1}%
\providecommand \href@noop [0]{\@secondoftwo}%
\providecommand \href [0]{\begingroup \@sanitize@url \@href}%
\providecommand \@href[1]{\@@startlink{#1}\@@href}%
\providecommand \@@href[1]{\endgroup#1\@@endlink}%
\providecommand \@sanitize@url [0]{\catcode `\\12\catcode `\$12\catcode `\&12\catcode `\#12\catcode `\^12\catcode `\_12\catcode `\%12\relax}%
\providecommand \@@startlink[1]{}%
\providecommand \@@endlink[0]{}%
\providecommand \url  [0]{\begingroup\@sanitize@url \@url }%
\providecommand \@url [1]{\endgroup\@href {#1}{\urlprefix }}%
\providecommand \urlprefix  [0]{URL }%
\providecommand \Eprint [0]{\href }%
\providecommand \doibase [0]{https://doi.org/}%
\providecommand \selectlanguage [0]{\@gobble}%
\providecommand \bibinfo  [0]{\@secondoftwo}%
\providecommand \bibfield  [0]{\@secondoftwo}%
\providecommand \translation [1]{[#1]}%
\providecommand \BibitemOpen [0]{}%
\providecommand \bibitemStop [0]{}%
\providecommand \bibitemNoStop [0]{.\EOS\space}%
\providecommand \EOS [0]{\spacefactor3000\relax}%
\providecommand \BibitemShut  [1]{\csname bibitem#1\endcsname}%
\let\auto@bib@innerbib\@empty
\bibitem [{\citenamefont {Stix}(1992)}]{Stix}%
  \BibitemOpen
  \bibfield  {author} {\bibinfo {author} {\bibfnamefont {T.~H.}\ \bibnamefont {Stix}},\ }\href@noop {} {\emph {\bibinfo {title} {Waves in plasmas}}}\ (\bibinfo  {publisher} {American Institute of Physics},\ \bibinfo {year} {1992})\BibitemShut {NoStop}%
\bibitem [{\citenamefont {Swanson}(2003)}]{Swanson}%
  \BibitemOpen
  \bibfield  {author} {\bibinfo {author} {\bibfnamefont {D.~G.}\ \bibnamefont {Swanson}},\ }\href@noop {} {\emph {\bibinfo {title} {Plasma Waves}}}\ (\bibinfo  {publisher} {Institute of Physics Publishing},\ \bibinfo {year} {2003})\BibitemShut {NoStop}%
\bibitem [{\citenamefont {Friedland}\ and\ \citenamefont {Bernstein}(1980)}]{Friedland}%
  \BibitemOpen
  \bibfield  {author} {\bibinfo {author} {\bibfnamefont {L.}~\bibnamefont {Friedland}}\ and\ \bibinfo {author} {\bibfnamefont {I.~B.}\ \bibnamefont {Bernstein}},\ }\bibfield  {title} {\enquote {\bibinfo {title} {General geometric optics formalism in plasmas},}\ }\href {https://doi.org/10.1109/TPS.1980.4317277} {\bibfield  {journal} {\bibinfo  {journal} {IEEE Trans. Plasma Sci.}\ }\textbf {\bibinfo {volume} {8}},\ \bibinfo {pages} {90--95} (\bibinfo {year} {1980})}\BibitemShut {NoStop}%
\bibitem [{\citenamefont {Tsironis}(2013)}]{Tsironis}%
  \BibitemOpen
  \bibfield  {author} {\bibinfo {author} {\bibfnamefont {C.}~\bibnamefont {Tsironis}},\ }\bibfield  {title} {\enquote {\bibinfo {title} {{On the Simplification of the Modeling of Electron-Cyclotron Wave Propagation in Thermonuclear Fusion Plasmas}},}\ }\href {https://doi.org/10.2528/PIERB12102911} {\bibfield  {journal} {\bibinfo  {journal} {Prog. Electromagn. Res. B}\ }\textbf {\bibinfo {volume} {47}},\ \bibinfo {pages} {37--61} (\bibinfo {year} {2013})}\BibitemShut {NoStop}%
\bibitem [{\citenamefont {Lau}\ \emph {et~al.}(2018)\citenamefont {Lau}, \citenamefont {Jaeger}, \citenamefont {Bertelli}, \citenamefont {Berry}, \citenamefont {Green}, \citenamefont {Murakami}, \citenamefont {Park}, \citenamefont {Pinsker},\ and\ \citenamefont {Prater}}]{Lau}%
  \BibitemOpen
  \bibfield  {author} {\bibinfo {author} {\bibfnamefont {C.}~\bibnamefont {Lau}}, \bibinfo {author} {\bibfnamefont {E.~F.}\ \bibnamefont {Jaeger}}, \bibinfo {author} {\bibfnamefont {N.}~\bibnamefont {Bertelli}}, \bibinfo {author} {\bibfnamefont {L.~A.}\ \bibnamefont {Berry}}, \bibinfo {author} {\bibfnamefont {D.~L.}\ \bibnamefont {Green}}, \bibinfo {author} {\bibfnamefont {M.}~\bibnamefont {Murakami}}, \bibinfo {author} {\bibfnamefont {J.~M.}\ \bibnamefont {Park}}, \bibinfo {author} {\bibfnamefont {R.~I.}\ \bibnamefont {Pinsker}},\ and\ \bibinfo {author} {\bibfnamefont {R.}~\bibnamefont {Prater}},\ }\bibfield  {title} {\enquote {\bibinfo {title} {{AORSA} full wave calculations of helicon waves in {DIII-D} and {ITER}},}\ }\href {https://doi.org/10.1088/1741-4326/aab96d} {\bibfield  {journal} {\bibinfo  {journal} {Nucl. Fusion}\ }\textbf {\bibinfo {volume} {58}},\ \bibinfo {pages} {066004} (\bibinfo {year} {2018})}\BibitemShut {NoStop}%
\bibitem [{\citenamefont {Wu}\ \emph {et~al.}(2021)\citenamefont {Wu}, \citenamefont {Bao}, \citenamefont {Cao}, \citenamefont {Chen}, \citenamefont {Chen}, \citenamefont {Chen}, \citenamefont {Chung}, \citenamefont {Deng}, \citenamefont {Du}, \citenamefont {Fan} \emph {et~al.}}]{Wu}%
  \BibitemOpen
  \bibfield  {author} {\bibinfo {author} {\bibfnamefont {Y.}~\bibnamefont {Wu}}, \bibinfo {author} {\bibfnamefont {W.-S.}\ \bibnamefont {Bao}}, \bibinfo {author} {\bibfnamefont {S.}~\bibnamefont {Cao}}, \bibinfo {author} {\bibfnamefont {F.}~\bibnamefont {Chen}}, \bibinfo {author} {\bibfnamefont {M.-C.}\ \bibnamefont {Chen}}, \bibinfo {author} {\bibfnamefont {X.}~\bibnamefont {Chen}}, \bibinfo {author} {\bibfnamefont {T.-H.}\ \bibnamefont {Chung}}, \bibinfo {author} {\bibfnamefont {H.}~\bibnamefont {Deng}}, \bibinfo {author} {\bibfnamefont {Y.}~\bibnamefont {Du}}, \bibinfo {author} {\bibfnamefont {D.}~\bibnamefont {Fan}}, \emph {et~al.},\ }\bibfield  {title} {\enquote {\bibinfo {title} {{Strong Quantum Computational Advantage Using a Superconducting Quantum Processor}},}\ }\href {https://doi.org/10.1103/PhysRevLett.127.180501} {\bibfield  {journal} {\bibinfo  {journal} {Phys. Rev. Lett.}\ }\textbf {\bibinfo {volume} {127}},\ \bibinfo {pages} {180501} (\bibinfo {year} {2021})}\BibitemShut {NoStop}%
\bibitem [{\citenamefont {Arute}\ \emph {et~al.}(2019)\citenamefont {Arute}, \citenamefont {Arya}, \citenamefont {Babbush}, \citenamefont {Bacon}, \citenamefont {Bardin}, \citenamefont {Barends}, \citenamefont {Biswas}, \citenamefont {Boixo}, \citenamefont {Brandao}, \citenamefont {Buell} \emph {et~al.}}]{Arute}%
  \BibitemOpen
  \bibfield  {author} {\bibinfo {author} {\bibfnamefont {F.}~\bibnamefont {Arute}}, \bibinfo {author} {\bibfnamefont {K.}~\bibnamefont {Arya}}, \bibinfo {author} {\bibfnamefont {R.}~\bibnamefont {Babbush}}, \bibinfo {author} {\bibfnamefont {D.}~\bibnamefont {Bacon}}, \bibinfo {author} {\bibfnamefont {J.~C.}\ \bibnamefont {Bardin}}, \bibinfo {author} {\bibfnamefont {R.}~\bibnamefont {Barends}}, \bibinfo {author} {\bibfnamefont {R.}~\bibnamefont {Biswas}}, \bibinfo {author} {\bibfnamefont {S.}~\bibnamefont {Boixo}}, \bibinfo {author} {\bibfnamefont {F.~G. S.~L.}\ \bibnamefont {Brandao}}, \bibinfo {author} {\bibfnamefont {D.~A.}\ \bibnamefont {Buell}}, \emph {et~al.},\ }\bibfield  {title} {\enquote {\bibinfo {title} {Quantum supremacy using a programmable superconducting processor},}\ }\href {https://doi.org/10.1038/s41586-019-1666-5} {\bibfield  {journal} {\bibinfo  {journal} {Nature}\ }\textbf {\bibinfo {volume} {574}},\ \bibinfo {pages} {505--510} (\bibinfo {year} {2019})}\BibitemShut {NoStop}%
\bibitem [{\citenamefont {Dodin}\ and\ \citenamefont {Startsev}(2021)}]{Dodin}%
  \BibitemOpen
  \bibfield  {author} {\bibinfo {author} {\bibfnamefont {I.~Y.}\ \bibnamefont {Dodin}}\ and\ \bibinfo {author} {\bibfnamefont {E.~A.}\ \bibnamefont {Startsev}},\ }\bibfield  {title} {\enquote {\bibinfo {title} {{On applications of quantum computing to plasma simulations}},}\ }\href {https://doi.org/10.1063/5.0056974} {\bibfield  {journal} {\bibinfo  {journal} {Phys. Plasmas}\ }\textbf {\bibinfo {volume} {28}},\ \bibinfo {pages} {092101} (\bibinfo {year} {2021})}\BibitemShut {NoStop}%
\bibitem [{\citenamefont {Joseph}\ \emph {et~al.}(2023)\citenamefont {Joseph}, \citenamefont {Shi}, \citenamefont {Porter}, \citenamefont {Castelli}, \citenamefont {Geyko}, \citenamefont {Graziani}, \citenamefont {Libby},\ and\ \citenamefont {DuBois}}]{Joseph}%
  \BibitemOpen
  \bibfield  {author} {\bibinfo {author} {\bibfnamefont {I.}~\bibnamefont {Joseph}}, \bibinfo {author} {\bibfnamefont {Y.}~\bibnamefont {Shi}}, \bibinfo {author} {\bibfnamefont {M.~D.}\ \bibnamefont {Porter}}, \bibinfo {author} {\bibfnamefont {A.~R.}\ \bibnamefont {Castelli}}, \bibinfo {author} {\bibfnamefont {V.~I.}\ \bibnamefont {Geyko}}, \bibinfo {author} {\bibfnamefont {F.~R.}\ \bibnamefont {Graziani}}, \bibinfo {author} {\bibfnamefont {S.~B.}\ \bibnamefont {Libby}},\ and\ \bibinfo {author} {\bibfnamefont {J.~L.}\ \bibnamefont {DuBois}},\ }\bibfield  {title} {\enquote {\bibinfo {title} {{Quantum computing for fusion energy science applications}},}\ }\href {https://doi.org/10.1063/5.0123765} {\bibfield  {journal} {\bibinfo  {journal} {Phys. Plasmas}\ }\textbf {\bibinfo {volume} {30}},\ \bibinfo {pages} {010501} (\bibinfo {year} {2023})}\BibitemShut {NoStop}%
\bibitem [{\citenamefont {Engel}, \citenamefont {Smith},\ and\ \citenamefont {Parker}(2019)}]{Engel}%
  \BibitemOpen
  \bibfield  {author} {\bibinfo {author} {\bibfnamefont {A.}~\bibnamefont {Engel}}, \bibinfo {author} {\bibfnamefont {G.}~\bibnamefont {Smith}},\ and\ \bibinfo {author} {\bibfnamefont {S.~E.}\ \bibnamefont {Parker}},\ }\bibfield  {title} {\enquote {\bibinfo {title} {Quantum algorithm for the vlasov equation},}\ }\href {https://doi.org/10.1103/PhysRevA.100.062315} {\bibfield  {journal} {\bibinfo  {journal} {Phys. Rev. A}\ }\textbf {\bibinfo {volume} {100}},\ \bibinfo {pages} {062315} (\bibinfo {year} {2019})}\BibitemShut {NoStop}%
\bibitem [{\citenamefont {Novikau}, \citenamefont {Startsev},\ and\ \citenamefont {Dodin}(2022)}]{Novikau}%
  \BibitemOpen
  \bibfield  {author} {\bibinfo {author} {\bibfnamefont {I.}~\bibnamefont {Novikau}}, \bibinfo {author} {\bibfnamefont {E.~A.}\ \bibnamefont {Startsev}},\ and\ \bibinfo {author} {\bibfnamefont {I.~Y.}\ \bibnamefont {Dodin}},\ }\bibfield  {title} {\enquote {\bibinfo {title} {Quantum signal processing for simulating cold plasma waves},}\ }\href {https://doi.org/10.1103/PhysRevA.105.062444} {\bibfield  {journal} {\bibinfo  {journal} {Phys. Rev. A}\ }\textbf {\bibinfo {volume} {105}},\ \bibinfo {pages} {062444} (\bibinfo {year} {2022})}\BibitemShut {NoStop}%
\bibitem [{\citenamefont {Low}\ and\ \citenamefont {Chuang}(2017)}]{Low}%
  \BibitemOpen
  \bibfield  {author} {\bibinfo {author} {\bibfnamefont {G.~H.}\ \bibnamefont {Low}}\ and\ \bibinfo {author} {\bibfnamefont {I.~L.}\ \bibnamefont {Chuang}},\ }\bibfield  {title} {\enquote {\bibinfo {title} {Optimal hamiltonian simulation by quantum signal processing},}\ }\href {https://doi.org/10.1103/PhysRevLett.118.010501} {\bibfield  {journal} {\bibinfo  {journal} {Phys. Rev. Lett.}\ }\textbf {\bibinfo {volume} {118}},\ \bibinfo {pages} {010501} (\bibinfo {year} {2017})}\BibitemShut {NoStop}%
\bibitem [{\citenamefont {Ameri}\ \emph {et~al.}(2023)\citenamefont {Ameri}, \citenamefont {Ye}, \citenamefont {Cappellaro}, \citenamefont {Krovi},\ and\ \citenamefont {Loureiro}}]{Ameri}%
  \BibitemOpen
  \bibfield  {author} {\bibinfo {author} {\bibfnamefont {A.}~\bibnamefont {Ameri}}, \bibinfo {author} {\bibfnamefont {E.}~\bibnamefont {Ye}}, \bibinfo {author} {\bibfnamefont {P.}~\bibnamefont {Cappellaro}}, \bibinfo {author} {\bibfnamefont {H.}~\bibnamefont {Krovi}},\ and\ \bibinfo {author} {\bibfnamefont {N.~F.}\ \bibnamefont {Loureiro}},\ }\bibfield  {title} {\enquote {\bibinfo {title} {Quantum algorithm for the linear vlasov equation with collisions},}\ }\href {https://doi.org/10.1103/PhysRevA.107.062412} {\bibfield  {journal} {\bibinfo  {journal} {Phys. Rev. A}\ }\textbf {\bibinfo {volume} {107}},\ \bibinfo {pages} {062412} (\bibinfo {year} {2023})}\BibitemShut {NoStop}%
\bibitem [{\citenamefont {Oscar Amaro}\ and\ \citenamefont {Cruz}(2023)}]{Amaro}%
  \BibitemOpen
  \bibfield  {author} {\bibinfo {author} {\bibnamefont {Oscar Amaro}}\ and\ \bibinfo {author} {\bibfnamefont {D.}~\bibnamefont {Cruz}},\ }\href@noop {} {\enquote {\bibinfo {title} {A living review of quantum computing for plasma physics},}\ } (\bibinfo {year} {2023}),\ \Eprint {https://arxiv.org/abs/2302.00001} {arXiv:2302.00001 [physics.plasm-ph]} \BibitemShut {NoStop}%
\bibitem [{\citenamefont {Koukoutsis}\ \emph {et~al.}(2023)\citenamefont {Koukoutsis}, \citenamefont {Hizanidis}, \citenamefont {Ram},\ and\ \citenamefont {Vahala}}]{Koukoutsis}%
  \BibitemOpen
  \bibfield  {author} {\bibinfo {author} {\bibfnamefont {E.}~\bibnamefont {Koukoutsis}}, \bibinfo {author} {\bibfnamefont {K.}~\bibnamefont {Hizanidis}}, \bibinfo {author} {\bibfnamefont {A.~K.}\ \bibnamefont {Ram}},\ and\ \bibinfo {author} {\bibfnamefont {G.}~\bibnamefont {Vahala}},\ }\bibfield  {title} {\enquote {\bibinfo {title} {{Dyson maps and unitary evolution for Maxwell equations in tensor dielectric media}},}\ }\href {https://doi.org/10.1103/PhysRevA.107.042215} {\bibfield  {journal} {\bibinfo  {journal} {Phys. Rev. A}\ }\textbf {\bibinfo {volume} {107}},\ \bibinfo {pages} {042215} (\bibinfo {year} {2023})}\BibitemShut {NoStop}%
\bibitem [{\citenamefont {Succi}, \citenamefont {Fillion-Gourdeau},\ and\ \citenamefont {Palpacelli}(2015)}]{Succi}%
  \BibitemOpen
  \bibfield  {author} {\bibinfo {author} {\bibfnamefont {S.}~\bibnamefont {Succi}}, \bibinfo {author} {\bibfnamefont {F.}~\bibnamefont {Fillion-Gourdeau}},\ and\ \bibinfo {author} {\bibfnamefont {S.}~\bibnamefont {Palpacelli}},\ }\bibfield  {title} {\enquote {\bibinfo {title} {{Quantum lattice Boltzmann is a quantum walk}},}\ }\href {https://doi.org/10.1140/epjqt/s40507-015-0025-1} {\bibfield  {journal} {\bibinfo  {journal} {EPJ Quantum Technol.}\ }\textbf {\bibinfo {volume} {2}},\ \bibinfo {pages} {12} (\bibinfo {year} {2015})}\BibitemShut {NoStop}%
\bibitem [{\citenamefont {Fillion-Gourdeau}, \citenamefont {MacLean},\ and\ \citenamefont {Laflamme}(2017)}]{Gourdeau}%
  \BibitemOpen
  \bibfield  {author} {\bibinfo {author} {\bibfnamefont {F.}\ \bibnamefont {Fillion-Gourdeau}}, \bibinfo {author} {\bibfnamefont {S.}~\bibnamefont {MacLean}},\ and\ \bibinfo {author} {\bibfnamefont {R.}~\bibnamefont {Laflamme}},\ }\bibfield  {title} {\enquote {\bibinfo {title} {Algorithm for the solution of the dirac equation on digital quantum computers},}\ }\href {https://doi.org/10.1103/PhysRevA.95.042343} {\bibfield  {journal} {\bibinfo  {journal} {Phys. Rev. A}\ }\textbf {\bibinfo {volume} {95}},\ \bibinfo {pages} {042343} (\bibinfo {year} {2017})}\BibitemShut {NoStop}%
\bibitem [{\citenamefont {Vahala}\ \emph {et~al.}(2023{\natexlab{a}})\citenamefont {Vahala}, \citenamefont {Soe}, \citenamefont {Koukoutsis}, \citenamefont {Hizanidis}, \citenamefont {Vahala},\ and\ \citenamefont {Ram}}]{Vahala3}%
  \BibitemOpen
  \bibfield  {author} {\bibinfo {author} {\bibfnamefont {G.}~\bibnamefont {Vahala}}, \bibinfo {author} {\bibfnamefont {M.}~\bibnamefont {Soe}}, \bibinfo {author} {\bibfnamefont {E.}~\bibnamefont {Koukoutsis}}, \bibinfo {author} {\bibfnamefont {K.}~\bibnamefont {Hizanidis}}, \bibinfo {author} {\bibfnamefont {L.}~\bibnamefont {Vahala}},\ and\ \bibinfo {author} {\bibfnamefont {A.~K.}\ \bibnamefont {Ram}},\ }\bibfield  {title} {\enquote {\bibinfo {title} {Qubit lattice algorithms based on the Schrodinger-Dirac representation of maxwell equations and their extensions},}\ }in\ \href {https://doi.org/10.5772/intechopen.112692} {\emph {\bibinfo {booktitle} {Schrodinger Equation - Fundamentals Aspects and Potential Applications}}},\ \bibinfo {editor} {edited by\ \bibinfo {editor} {\bibfnamefont {D.~M.~B.}\ \bibnamefont {Tahir}}, \bibinfo {editor} {\bibfnamefont {D.~M.}\ \bibnamefont {Sagir}}, \bibinfo {editor} {\bibfnamefont {A.~P. M.~I.}\ \bibnamefont {Khan}}, \bibinfo {editor} {\bibfnamefont {D.~M.}\ \bibnamefont
  {Rafique}},\ and\ \bibinfo {editor} {\bibfnamefont {D.~F.}\ \bibnamefont {Bulnes}}}\ (\bibinfo  {publisher} {IntechOpen},\ \bibinfo {address} {Rijeka},\ \bibinfo {year} {2023})\ Chap.~\bibinfo {chapter} {5}\BibitemShut {NoStop}%
\bibitem [{\citenamefont {Vahala}\ \emph {et~al.}(2021)\citenamefont {Vahala}, \citenamefont {Vahala}, \citenamefont {Soe},\ and\ \citenamefont {Ram}}]{Vahala4}%
  \BibitemOpen
  \bibfield  {author} {\bibinfo {author} {\bibfnamefont {G.}~\bibnamefont {Vahala}}, \bibinfo {author} {\bibfnamefont {L.}~\bibnamefont {Vahala}}, \bibinfo {author} {\bibfnamefont {M.}~\bibnamefont {Soe}},\ and\ \bibinfo {author} {\bibfnamefont {A.~K.}\ \bibnamefont {Ram}},\ }\bibfield  {title} {\enquote {\bibinfo {title} {One- and two-dimensional quantum lattice algorithms for Maxwell equations in inhomogeneous scalar dielectric media I: theory},}\ }\href {https://doi.org/10.1080/10420150.2021.1891058} {\bibfield  {journal} {\bibinfo  {journal} {Radiat. Eff. Defects Solids}\ }\textbf {\bibinfo {volume} {176}},\ \bibinfo {pages} {49--63} (\bibinfo {year} {2021})}\BibitemShut {NoStop}%
\bibitem [{\citenamefont {Vahala}\ \emph {et~al.}(2020)\citenamefont {Vahala}, \citenamefont {Vahala}, \citenamefont {Soe},\ and\ \citenamefont {Ram}}]{Vahala1}%
  \BibitemOpen
  \bibfield  {author} {\bibinfo {author} {\bibfnamefont {G.}~\bibnamefont {Vahala}}, \bibinfo {author} {\bibfnamefont {L.}~\bibnamefont {Vahala}}, \bibinfo {author} {\bibfnamefont {M.}~\bibnamefont {Soe}},\ and\ \bibinfo {author} {\bibfnamefont {A.~K.}\ \bibnamefont {Ram}},\ }\bibfield  {title} {\enquote {\bibinfo {title} {Unitary quantum lattice simulations for Maxwell equations in vacuum and in dielectric media},}\ }\href {https://doi.org/10.1017/S0022377820001166} {\bibfield  {journal} {\bibinfo  {journal} {J. Plasma Phys.}\ }\textbf {\bibinfo {volume} {86}},\ \bibinfo {pages} {905860518} (\bibinfo {year} {2020})}\BibitemShut {NoStop}%
\bibitem [{\citenamefont {Vahala}\ \emph {et~al.}(2022)\citenamefont {Vahala}, \citenamefont {Hawthorne}, \citenamefont {Vahala}, \citenamefont {Ram},\ and\ \citenamefont {Soe}}]{Vahala2}%
  \BibitemOpen
  \bibfield  {author} {\bibinfo {author} {\bibfnamefont {G.}~\bibnamefont {Vahala}}, \bibinfo {author} {\bibfnamefont {J.}~\bibnamefont {Hawthorne}}, \bibinfo {author} {\bibfnamefont {L.}~\bibnamefont {Vahala}}, \bibinfo {author} {\bibfnamefont {A.~K.}\ \bibnamefont {Ram}},\ and\ \bibinfo {author} {\bibfnamefont {M.}~\bibnamefont {Soe}},\ }\bibfield  {title} {\enquote {\bibinfo {title} {Quantum lattice representation for the curl equations of Maxwell equations},}\ }\href {https://doi.org/10.1080/10420150.2022.2049784} {\bibfield  {journal} {\bibinfo  {journal} {Radiat. Eff. Defects Solids}\ }\textbf {\bibinfo {volume} {177}},\ \bibinfo {pages} {85--94} (\bibinfo {year} {2022})}\BibitemShut {NoStop}%
\bibitem [{\citenamefont {Ram}\ \emph {et~al.}(2021)\citenamefont {Ram}, \citenamefont {Vahala}, \citenamefont {Vahala},\ and\ \citenamefont {Soe}}]{Ram}%
  \BibitemOpen
  \bibfield  {author} {\bibinfo {author} {\bibfnamefont {A.~K.}\ \bibnamefont {Ram}}, \bibinfo {author} {\bibfnamefont {G.}~\bibnamefont {Vahala}}, \bibinfo {author} {\bibfnamefont {L.}~\bibnamefont {Vahala}},\ and\ \bibinfo {author} {\bibfnamefont {M.}~\bibnamefont {Soe}},\ }\bibfield  {title} {\enquote {\bibinfo {title} {{Reflection and transmission of electromagnetic pulses at a planar dielectric interface: Theory and quantum lattice simulations}},}\ }\href {https://doi.org/10.1063/5.0067204} {\bibfield  {journal} {\bibinfo  {journal} {AIP Advance}\ }\textbf {\bibinfo {volume} {11}},\ \bibinfo {pages} {105116} (\bibinfo {year} {2021})}\BibitemShut {NoStop}%
\bibitem [{\citenamefont {Vahala}\ \emph {et~al.}(2023)\citenamefont {Vahala}, \citenamefont {Soe}, \citenamefont {Vahala}, \citenamefont {Ram}, \citenamefont {Koukoutsis},\ and\ \citenamefont {Hizanidis}}]{unpublished}%
  \BibitemOpen
  \bibfield  {author} {\bibinfo {author} {\bibfnamefont {G.}~\bibnamefont {Vahala}}, \bibinfo {author} {\bibfnamefont {M.}~\bibnamefont {Soe}}, \bibinfo {author} {\bibfnamefont {L.}~\bibnamefont {Vahala}}, \bibinfo {author} {\bibfnamefont {A.~K.}\ \bibnamefont {Ram}}, \bibinfo {author} {\bibfnamefont {E.}~\bibnamefont {Koukoutsis}},\ and\ \bibinfo {author} {\bibfnamefont {K.}~\bibnamefont {Hizanidis}},\ }\bibfield  {title} {\enquote {\bibinfo {title} {Qubit lattice algorithm simulations of Maxwell's equations for scattering from anisotropic dielectric objects},}\ }\href {https://doi.org/10.1016/j.compfluid.2023.106039} {\bibfield  {journal} {\bibinfo  {journal} {Comput. Fluids}\ }\textbf {\bibinfo {volume} {266}},\ \bibinfo {pages} {106039} (\bibinfo {year} {2023})}\BibitemShut {NoStop}%
\bibitem [{\citenamefont {Vahala}, \citenamefont {Yepez},\ and\ \citenamefont {Vahala}(2003)}]{Vahala5}%
  \BibitemOpen
  \bibfield  {author} {\bibinfo {author} {\bibfnamefont {G.}~\bibnamefont {Vahala}}, \bibinfo {author} {\bibfnamefont {J.}~\bibnamefont {Yepez}},\ and\ \bibinfo {author} {\bibfnamefont {L.}~\bibnamefont {Vahala}},\ }\bibfield  {title} {\enquote {\bibinfo {title} {Quantum lattice gas representation of some classical solitons},}\ }\href {https://doi.org/10.1016/S0375-9601(03)00334-7} {\bibfield  {journal} {\bibinfo  {journal} {Physics Letters A}\ }\textbf {\bibinfo {volume} {310}},\ \bibinfo {pages} {187--196} (\bibinfo {year} {2003})}\BibitemShut {NoStop}%
\bibitem [{\citenamefont {Vahala}\ \emph {et~al.}(2019)\citenamefont {Vahala}, \citenamefont {Vahala}, \citenamefont {Soe}, \citenamefont {Ram},\ and\ \citenamefont {Yepez}}]{Linda1}%
  \BibitemOpen
  \bibfield  {author} {\bibinfo {author} {\bibfnamefont {L.}~\bibnamefont {Vahala}}, \bibinfo {author} {\bibfnamefont {G.}~\bibnamefont {Vahala}}, \bibinfo {author} {\bibfnamefont {M.}~\bibnamefont {Soe}}, \bibinfo {author} {\bibfnamefont {A.}~\bibnamefont {Ram}},\ and\ \bibinfo {author} {\bibfnamefont {J.}~\bibnamefont {Yepez}},\ }\bibfield  {title} {\enquote {\bibinfo {title} {Unitary qubit lattice algorithm for three-dimensional vortex solitons in hyperbolic self-defocusing media},}\ }\href {https://doi.org/10.1016/j.cnsns.2019.03.016} {\bibfield  {journal} {\bibinfo  {journal} {Commun. Nonlinear Sci. Numer. Simul.}\ }\textbf {\bibinfo {volume} {75}},\ \bibinfo {pages} {152--159} (\bibinfo {year} {2019})}\BibitemShut {NoStop}%
\bibitem [{\citenamefont {Boghosian}\ and\ \citenamefont {Taylor}(1998)}]{Boghosian}%
  \BibitemOpen
  \bibfield  {author} {\bibinfo {author} {\bibfnamefont {B.~M.}\ \bibnamefont {Boghosian}}\ and\ \bibinfo {author} {\bibfnamefont {W.}~\bibnamefont {Taylor}},\ }\bibfield  {title} {\enquote {\bibinfo {title} {Simulating quantum mechanics on a quantum computer},}\ }\href {https://doi.org/10.1016/S0167-2789(98)00042-6} {\bibfield  {journal} {\bibinfo  {journal} {Physica D}\ }\textbf {\bibinfo {volume} {120}},\ \bibinfo {pages} {30--42} (\bibinfo {year} {1998})}\BibitemShut {NoStop}%
\bibitem [{\citenamefont {Yepez}(2002{\natexlab{a}})}]{Yepez}%
  \BibitemOpen
  \bibfield  {author} {\bibinfo {author} {\bibfnamefont {J.}~\bibnamefont {Yepez}}\ and\ \bibinfo {author} {\bibfnamefont {B.}~\bibnamefont {Boghosian}},\ }\bibfield  {title} {\enquote {\bibinfo {title} {An efficient and accurate quantum lattice-gas model for the many-body Schrodinger wave equation},}\ }\href {https://doi.org/10.1016/S0010-4655(02)00419-8} {\bibfield  {journal} {\bibinfo  {journal} {Comput. Phys. Commun.}\ }\textbf {\bibinfo {volume} {146}},\ \bibinfo {pages} {280--294} (\bibinfo {year} {2002}{\natexlab{a}})}\BibitemShut {NoStop}%
\bibitem [{\citenamefont {Yepez}(2002{\natexlab{b}})}]{Yepez2}%
  \BibitemOpen
  \bibfield  {author} {\bibinfo {author} {\bibfnamefont {J.}~\bibnamefont {Yepez}},\ }\href@noop {} {\enquote {\bibinfo {title} {An efficient and accurate quantum algorithm for the dirac equation},}\ } (\bibinfo {year} {2002}{\natexlab{b}}),\ \Eprint {https://arxiv.org/abs/quant-ph/0210093} {arXiv:quant-ph/0210093 [quant-ph]} \BibitemShut {NoStop}%
\bibitem [{\citenamefont {Silveirinha}(2015)}]{Silveirinha}%
  \BibitemOpen
  \bibfield  {author} {\bibinfo {author} {\bibfnamefont {M.~G.}\ \bibnamefont {Silveirinha}},\ }\bibfield  {title} {\enquote {\bibinfo {title} {Chern invariants for continuous media},}\ }\href {https://doi.org/10.1103/PhysRevB.92.125153} {\bibfield  {journal} {\bibinfo  {journal} {Phys. Rev. B}\ }\textbf {\bibinfo {volume} {92}},\ \bibinfo {pages} {125153} (\bibinfo {year} {2015})}\BibitemShut {NoStop}%
\bibitem [{\citenamefont {Cassier}, \citenamefont {Joly},\ and\ \citenamefont {Kachanovska}(2017)}]{Cassier}%
  \BibitemOpen
  \bibfield  {author} {\bibinfo {author} {\bibfnamefont {M.}~\bibnamefont {Cassier}}, \bibinfo {author} {\bibfnamefont {P.}~\bibnamefont {Joly}},\ and\ \bibinfo {author} {\bibfnamefont {M.}~\bibnamefont {Kachanovska}},\ }\bibfield  {title} {\enquote {\bibinfo {title} {Mathematical models for dispersive electromagnetic waves: An overview},}\ }\href {https://doi.org/https://doi.org/10.1016/j.camwa.2017.07.025} {\bibfield  {journal} {\bibinfo  {journal} {Comput. Math. with Appl.}\ }\textbf {\bibinfo {volume} {74}},\ \bibinfo {pages} {2792--2830} (\bibinfo {year} {2017})}\BibitemShut {NoStop}%
\bibitem [{\citenamefont {Lee}\ and\ \citenamefont {Kalluri}(1999)}]{Lee}%
  \BibitemOpen
  \bibfield  {author} {\bibinfo {author} {\bibfnamefont {J.~H.}\ \bibnamefont {Lee}}\ and\ \bibinfo {author} {\bibfnamefont {D.~K.}\ \bibnamefont {Kalluri}},\ }\bibfield  {title} {\enquote {\bibinfo {title} {Three-dimensional fdtd simulation of electromagnetic wave transformation in a dynamic inhomogeneous magnetized plasma},}\ }\href {https://doi.org/10.1109/8.785745} {\bibfield  {journal} {\bibinfo  {journal} {IEEE Trans. Antennas Propag.}\ }\textbf {\bibinfo {volume} {47}},\ \bibinfo {pages} {1146--1151} (\bibinfo {year} {1999})}\BibitemShut {NoStop}%
\bibitem [{\citenamefont {Khan}(2005)}]{Khan}%
  \BibitemOpen
  \bibfield  {author} {\bibinfo {author} {\bibfnamefont {S.~A.}\ \bibnamefont {Khan}},\ }\bibfield  {title} {\enquote {\bibinfo {title} {{An Exact Matrix Representation of Maxwell's Equations}},}\ }\href {https://doi.org/10.1238/Physica.Regular.071a00440} {\bibfield  {journal} {\bibinfo  {journal} {Phys. Scr.}\ }\textbf {\bibinfo {volume} {71}},\ \bibinfo {pages} {440} (\bibinfo {year} {2005})}\BibitemShut {NoStop}%
\bibitem [{\citenamefont {Novikau}, \citenamefont {Dodin},\ and\ \citenamefont {Startsev}(2023)}]{Novikau2}%
  \BibitemOpen
  \bibfield  {author} {\bibinfo {author} {\bibfnamefont {I.}~\bibnamefont {Novikau}}, \bibinfo {author} {\bibfnamefont {I.}~\bibnamefont {Dodin}},\ and\ \bibinfo {author} {\bibfnamefont {E.}~\bibnamefont {Startsev}},\ }\bibfield  {title} {\enquote {\bibinfo {title} {Simulation of linear non-hermitian boundary-value problems with quantum singular-value transformation},}\ }\href {https://doi.org/10.1103/PhysRevApplied.19.054012} {\bibfield  {journal} {\bibinfo  {journal} {Phys. Rev. Appl.}\ }\textbf {\bibinfo {volume} {19}},\ \bibinfo {pages} {054012} (\bibinfo {year} {2023})}\BibitemShut {NoStop}%
\bibitem [{\citenamefont {Bullock}\ and\ \citenamefont {Markov}(2004)}]{Bullock}%
  \BibitemOpen
  \bibfield  {author} {\bibinfo {author} {\bibfnamefont {S.~S.}\ \bibnamefont {Bullock}}\ and\ \bibinfo {author} {\bibfnamefont {I.~L.}\ \bibnamefont {Markov}},\ }\bibfield  {title} {\enquote {\bibinfo {title} {Asymptotically optimal circuits for arbitrary n-qubit diagonal comutations},}\ }\href {https://doi.org/10.5555/2011572.2011575} {\bibfield  {journal} {\bibinfo  {journal} {Quantum Inf. Comput.}\ }\textbf {\bibinfo {volume} {4}},\ \bibinfo {pages} {27--47} (\bibinfo {year} {2004})}\BibitemShut {NoStop}%
\bibitem [{\citenamefont {Childs}, \citenamefont {Kothari},\ and\ \citenamefont {Somma}(2017)}]{Childs}%
  \BibitemOpen
  \bibfield  {author} {\bibinfo {author} {\bibfnamefont {A.~M.}\ \bibnamefont {Childs}}, \bibinfo {author} {\bibfnamefont {R.}~\bibnamefont {Kothari}},\ and\ \bibinfo {author} {\bibfnamefont {R.~D.}\ \bibnamefont {Somma}},\ }\bibfield  {title} {\enquote {\bibinfo {title} {{Quantum Algorithm for Systems of Linear Equations with Exponentially Improved Dependence on Precision}},}\ }\href {https://doi.org/10.1137/16M1087072} {\bibfield  {journal} {\bibinfo  {journal} {SIAM J. Comput.}\ }\textbf {\bibinfo {volume} {46}},\ \bibinfo {pages} {1920--1950} (\bibinfo {year} {2017})}\BibitemShut {NoStop}%
\bibitem [{\citenamefont {Taylor}, \citenamefont {Smith},\ and\ \citenamefont {Yepez}(2019)}]{Taylor}%
  \BibitemOpen
  \bibfield  {author} {\bibinfo {author} {\bibfnamefont {J.}~\bibnamefont {Taylor}}, \bibinfo {author} {\bibfnamefont {S.}~\bibnamefont {Smith}},\ and\ \bibinfo {author} {\bibfnamefont {J.}~\bibnamefont {Yepez}},\ }\href@noop {} {\enquote {\bibinfo {title} {Spin-2 bec spinor superfluid soliton-soliton scattering in one and two space dimensions},}\ } (\bibinfo {year} {2019}),\ \Eprint {https://arxiv.org/abs/1907.12834} {arXiv:1907.12834 [cond-mat.quant-gas]} \BibitemShut {NoStop}%
\bibitem [{\citenamefont {Ram}\ and\ \citenamefont {Hizanidis}(2016)}]{Ram2}%
  \BibitemOpen
  \bibfield  {author} {\bibinfo {author} {\bibfnamefont {A.~K.}\ \bibnamefont {Ram}}\ and\ \bibinfo {author} {\bibfnamefont {K.}~\bibnamefont {Hizanidis}},\ }\bibfield  {title} {\enquote {\bibinfo {title} {{Scattering of radio frequency waves by cylindrical density filaments in tokamak plasmas}},}\ }\href {https://doi.org/10.1063/1.4941588} {\bibfield  {journal} {\bibinfo  {journal} {Phys. Plasmas}\ }\textbf {\bibinfo {volume} {23}},\ \bibinfo {pages} {022504} (\bibinfo {year} {2016})}\BibitemShut {NoStop}%
\bibitem [{\citenamefont {Ram}, \citenamefont {Hizanidis},\ and\ \citenamefont {Kominis}(2013)}]{Ram3}%
  \BibitemOpen
  \bibfield  {author} {\bibinfo {author} {\bibfnamefont {A.~K.}\ \bibnamefont {Ram}}, \bibinfo {author} {\bibfnamefont {K.}~\bibnamefont {Hizanidis}},\ and\ \bibinfo {author} {\bibfnamefont {Y.}~\bibnamefont {Kominis}},\ }\bibfield  {title} {\enquote {\bibinfo {title} {{Scattering of radio frequency waves by blobs in tokamak plasmas}},}\ }\href {https://doi.org/10.1063/1.4803898} {\bibfield  {journal} {\bibinfo  {journal} {Phys. Plasmas}\ }\textbf {\bibinfo {volume} {20}},\ \bibinfo {pages} {056110} (\bibinfo {year} {2013})}\BibitemShut {NoStop}%
\bibitem [{\citenamefont {Jackson}(1998)}]{Jackson}%
  \BibitemOpen
  \bibfield  {author} {\bibinfo {author} {\bibfnamefont {J.~D.}\ \bibnamefont {Jackson}},\ }\href@noop {} {\emph {\bibinfo {title} {Classical Electrodynamics}}}\ (\bibinfo  {publisher} {Wiley},\ \bibinfo {year} {1998})\BibitemShut {NoStop}%
\end{thebibliography}
\end{document}